# On Interfacing the Brain with Quantum Computers: An Approach to Listen to the Logic of the Mind

Eduardo R. Miranda
Interdisciplinary Centre for Computer Music Research (ICCMR)
University of Plymouth, United Kingdom
eduardo.miranda@plymouth.ac.uk

## 1 Introduction

The human brain is allegedly the most complex object known to mankind: it has circa one hundred billion neurones forming a network of quadrillions of connections. The amount of information that circulates through this network is immense.

There is a school of thought, called Dualism, which considers the mind and the brain as separate entities (Rozemond, 1988). What is more, it has been suggested that minds would not even need brains to exist. Although the separation between mind and brain enjoys some currency in philosophical circles, it is generally agreed nowadays that the mind results from the functioning of the brain.

The scientific community does not have yet a clear understanding of how brain activity gives rise to the mind. Even though the behaviour of individual neurones is fairly well understood nowadays, the way in which they cooperate in ensembles of millions has been difficult to unpack. Yet, this is of paramount importance to fathom how the brain creates the mind.

The advent of increasingly sophisticated brain scanning technology has been enabling a plethora of research activity to comprehend the neuronal correlates of mental activities (Kraft et al., 2008; Vartanian et al., 2013). This comprehension is paramount for philosophy, psychology, medicine and engineering. Indeed, emerging technology that enables users to control systems with their mind banks on such developments.

A Brain-Computer Interface, or BCI, is a piece of equipment that enables users to control systems with their mind. It reads signals from the brain and harness them for communicating with a computer, or controlling a mechanical device, such as a robotic arm, a musical instrument or a wheelchair (Figure 1).

This chapter presents an approach to study and harness neuronal correlates of mental activity for the development of BCI systems. It introduces the notion of a *logic of the mind*, where neurophysiological data are encoded as logical expressions representing mental activity. The logic of the mind is proposed as a tool for studying the neuronal correlates of mental activities. And these logical expressions can be associated with commands for a BCI system.

Effective logical expressions are likely to be extensive, involving dozens of variables. Large expressions require considerable computational power to be processed. This is problematic for BCI applications because they require fast reaction times to execute sequences of commands.

Quantum computers hold much promise in terms of processing speed for some problems, including those involving logical expressions. Hence the desire to use quantum computers to process the logic of the mind.

Quantum computers are fundamentally different from the typical computer as we know it. The speed-up prophecy depends, amongst other things, on algorithms that cleverly exploit fundamental properties of quantum physics, such as *superpositio*n, *entanglement* and *interference*.

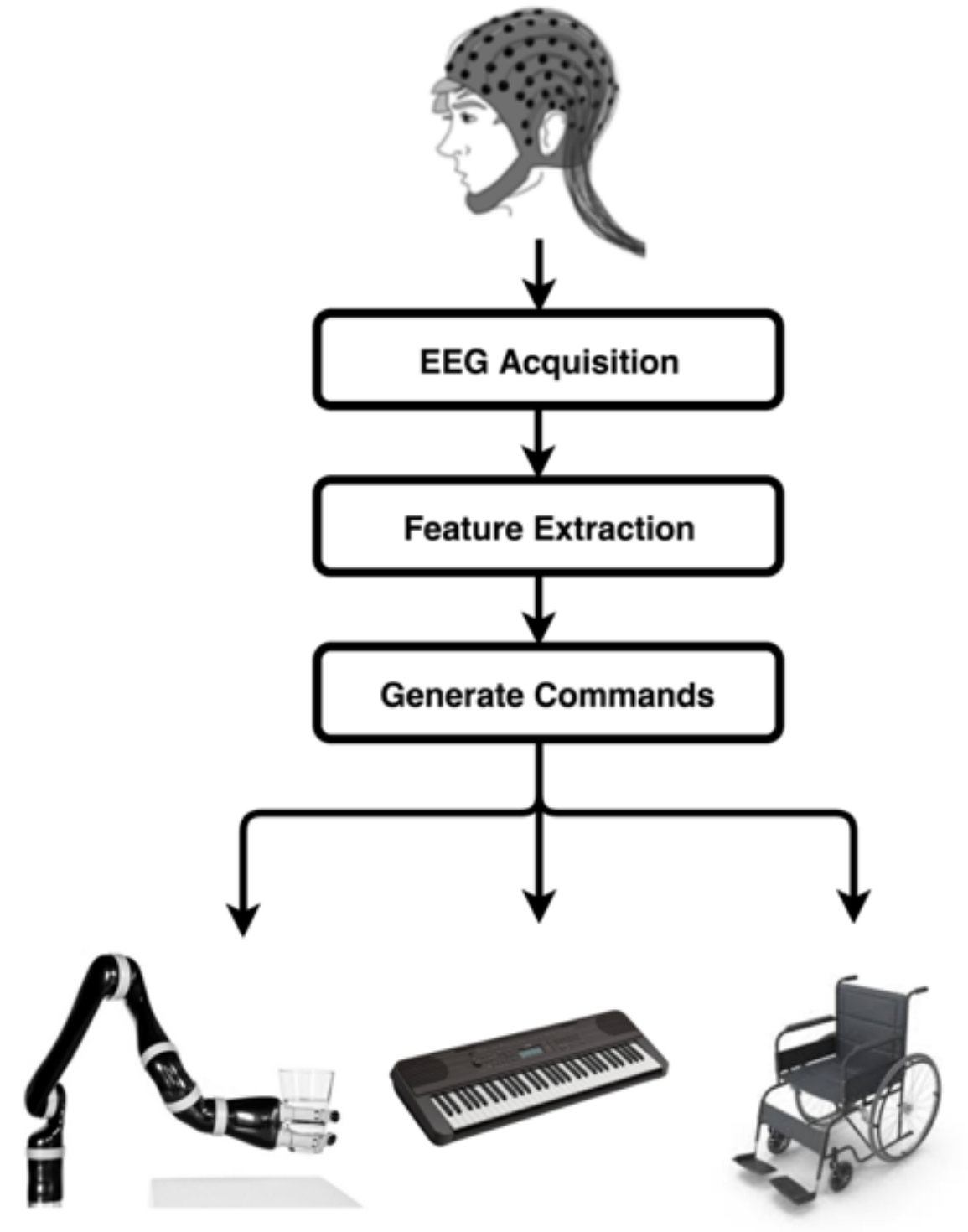

**Figure 1:** A BCI harness brain signals to communicate with devices.

The chapter begins with an introduction to BCI and the electroencephalogram, which is the neurophysiological signal that is normally used in BCI. Then, it briefly discusses how the EEG corresponds to mental states, followed by an introduction to the logic of the mind. After that, there is an overview of quantum computing, focusing on the basics deemed necessary to understand how it processes logical expressions.

An example of a BCI is presented to illustrate how the concepts discussed thus far fit together in practice. In a nutshell, the system reads the EEG and builds logical expressions, which are sent to a quantum computer to solve them. In turn, the system converts the results into sounds by means of a bespoke synthesiser. Essentially, the BCI here is a musical instrument controlled by the mind of the player.

Our BCI is a proof-of-concept aimed at demonstrating how quantum computing may support the development of sophisticated BCI systems. The remaining of the chapter is devoted to technical and practical considerations on the limitations of current quantum computing hardware technology and scalability of the system.

## 2 Brain-Computer Interfacing

By and large, BCI research is concerned with the development of assistive technology for persons with severe motor impairment. However, BCI technology is also being developed for other applications, such as computer games (Hasan and Gan, 2012), biometrics (Palaniappan, 2008) and cursor control (Wilson and Palaniappan, 2011). And this author has developed BCI for making music (Miranda, 2006; Miranda et al., 2011).

There are different types of brain signals and respective sensors to read them. The type of signal that is most commonly used for BCI technology is the electroencephalogram.

### 2.1 The electroencephalogram

The most common brain cells are neurones and glia. Neurones communicate with one another through electrical impulses. Neuronal electrical activity can be recorded with electrodes placed on the scalp. This recording is called the electroencephalogram, or EEG. The EEG conveys the overall activity of millions of neurones in the brain in terms of electric current. This is measured as the voltage difference between two or more electrodes, one of which is taken as a reference. See (Marcuse et al., 2015) for more details.

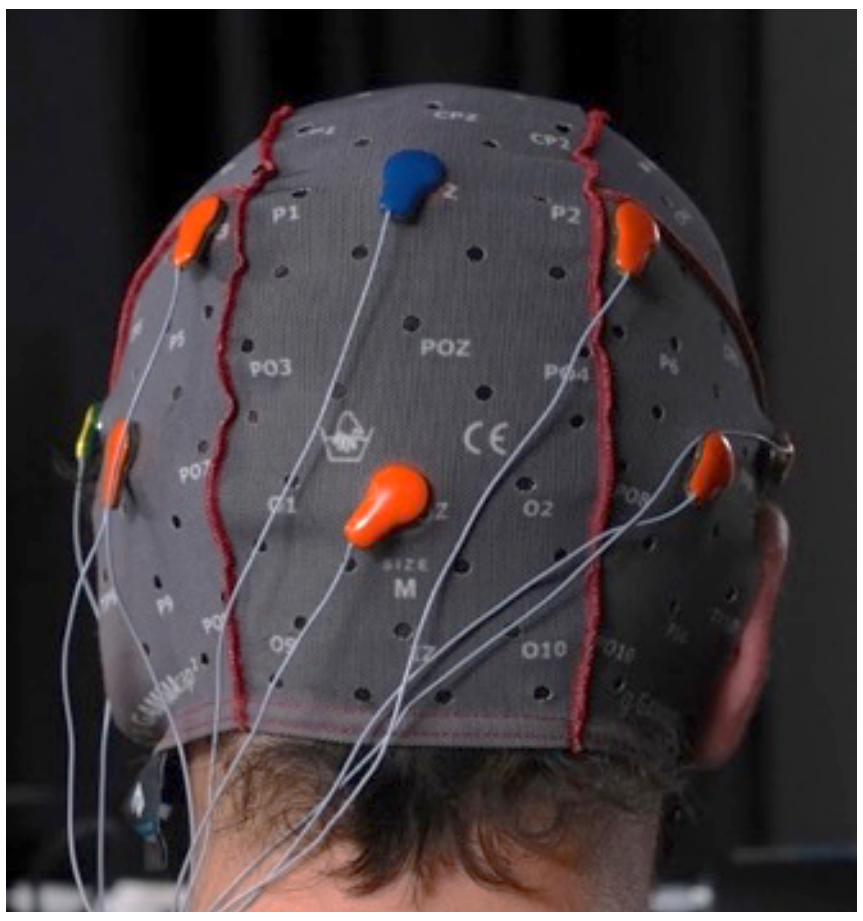

**Figure 2:** Cap furnished with electrodes relay EEG data to a computer for further processing.

It is also possible to record electrical brain activity with electrodes surgically implanted under the skull, on the surface of the cortex or deep inside the brain. This is often called electrocorticography (ECoG) or intracranial EEG (iEEG). Whereas implanted electrodes provide a far better signal to work with than surface ones, brain implants are not yet routinely used for BCI systems for health and safety reasons. Other technologies for brain scanning include functional Magnetic Resonance Imaging (fMRI), near-infrared spectroscopy (NIRS) and magnetoencephalography (MEG).

However, these technologies are prohibitively expensive, less portable, and/or offer inadequate time-resolution resolution for a BCI application (McFarland and Wolpaw, 2017).

For this project, we used a device manufactured by g.tec[1] to read the EEG. It consists of a cap furnished with electrodes and a transmitter that relays the EEG wirelessly to a computer for further processing (Figure 2)

Electrodes positioning on the head may vary depending on the purpose of the system or experiment. A commonly adopted convention is shown in Figure 3. The terminology for referring to the positioning of electrodes uses letters to indicate a brain region and a number: Fp (for pre-frontal), F (for frontal), C (for central), T (for temporal), P (p for parietal) and O (for occipital). Odd numbers are for electrodes on the left side of the head and even numbers are for those on the right side; the letter ‘z’ stands for the central region.

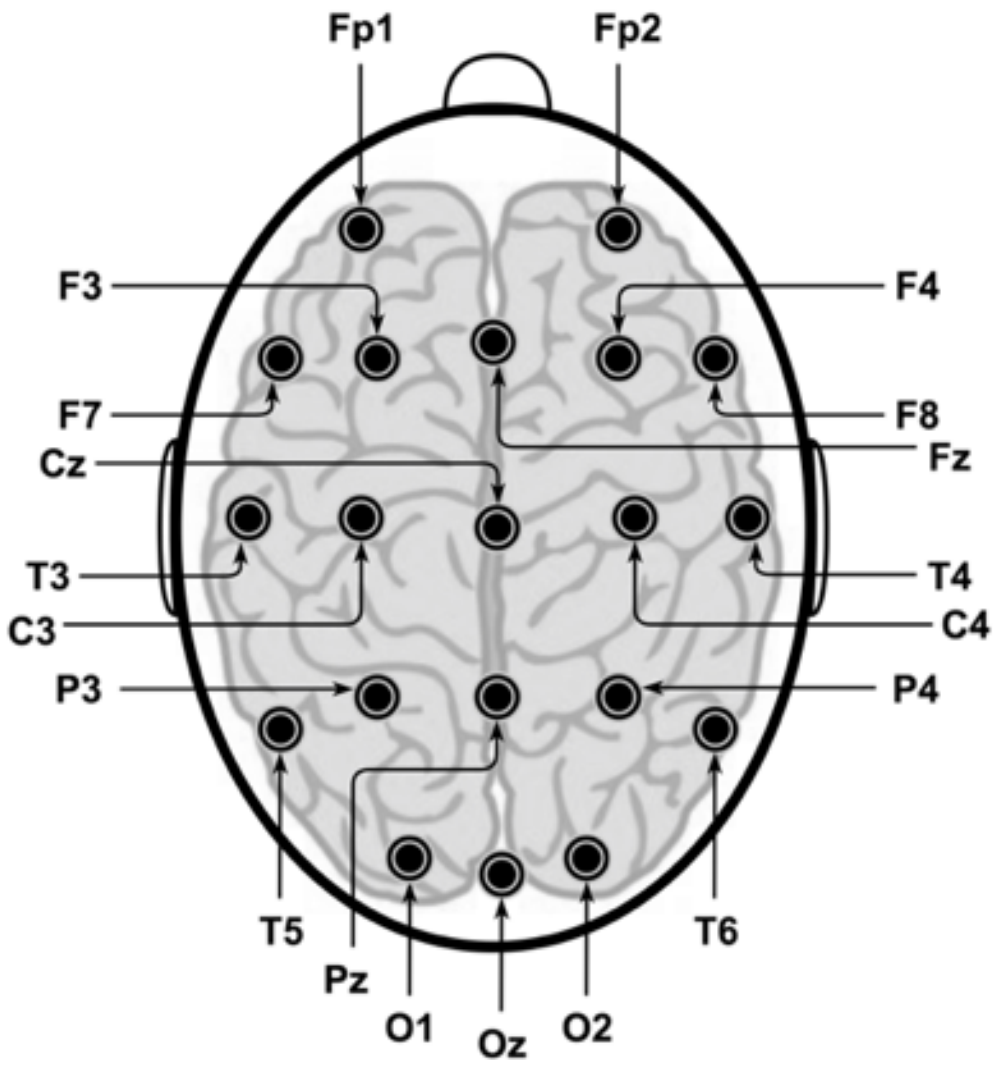

**Figure 3:** Convention for placing electrodes on the scalp.

The EEG from scalp electrodes is a difficult signal to handle. It is filtered by the meninges (the membranes that separate the cortex from the skull), the skull and the skin before it reaches the electrodes. Moreover, the signal is weak: it is measured in terms of microvolts ($\mu V$). It needs to be amplified considerably in order to be useful for a BCI. But amplification invariably brings spurious signals. Thus, the EEG needs to be harnessed with signal processing and analysis methods in order to render it suitable for a BCI system.

Power spectrum analysis is a popular method to harness the EEG. This method breaks the EEG signal into different frequency bands and reveals the distribution of power between them. Power spectrum analysis is useful because it can reveal patterns of brain activity, and a computer can be programmed to recognise and translate into commands for a system. Although this chapter focuses on power spectrum analysis, it is worth noting that there are other EEG analysis methods for detecting mental activity as well; e.g., Hjorth parameters (Oh et al., 2014).

[1] https://www.gtec.at/

Typically, users must learn how to voluntarily produce specific patterns of EEG signals in order to be able to control something with a BCI. A fair amount of research is underway to develop lexicons of detectable EEG patterns, understand what they mean, and develop methods to train users to produce them.

### 2.2 The semantics of the EEG

The EEG conveys information about mental activity (Anderson and Sijercic, 1996; Petsche and Etlinger, 1998). In medicine, the EEG is an important tool for diagnosis of brain disorders.

There has been an increasing amount of research aimed at the identification of EEG correlates of all sorts of mental activities (Nicolas-Alonso and Gomez-Gil, 2012; Hinterberger et al., 2014; So et al., 2017; Yelamanchili, 2018; Jeunet, 2020). For instance, Giannitrapani (1985) identified EEG patterns associated with abstract intellectual tasks, such as doing arithmetic operations mentally. Guenter and Brumberg (2011) presented neural correlates for the production of speech, which can be detected in the EEG. And more recently, Daly et al. (2018) detected EEG signatures correlated with emotions elicited while subjects were listening to music.

It is generally known that the distribution of power in the spectrum of the EEG can indicate different states of mind. For example, a spectrum with salient low-frequency components is often associated with a state of drowsiness, whereas a spectrum with salient high-frequencies is associated with a state of alertness. Giannitrapani (1985) linked the prominence of low-frequency components with a passive state of mind, as opposed to an active state, which is characterised by high-frequency spectral components.

Research exploring mental correlates of the EEG normally considers spectral components up to 40 Hz (Kropotov, 2008). There are four recognised spectral frequency bands, also known as EEG rhythms, each of which is associated with specific mental states (Table 1).

| Bands | Rhythms | Mental States |
|---|---|---|
| $f < 4$ | delta | Sleep; can indicate of cerebral anomaly. |
| $4 \leq f < 8$ | theta | Drowsiness; hypnotic state; can indicate cerebral anomaly. |
| $8 \leq f < 15$ | alpha | Relaxed, meditative, unfocused, almost drowsy state of mind. |
| $15 \leq f < 40$ | beta | High arousal, active thinking, consciously focusing state of mind. |

**Table 1:** Typical EEG rhythms and associated mental states. Frequency bands are in Hertz (Hz).

The exact boundaries of the bands listed in Table 1 and their respective mental states vary from one author to another. And the EEG that gives raise to mental states is highly dynamic.

Figure 4 shows a snapshot of EEG mapped onto 2D and 3D topographic representations using the *OpenVibe* software (Renard et al., 2010). The line cursor on the ‘Signal display’ pane shows the precise moment of the snapshot. At this moment, prominent beta rhythms are detected at the lateral sides of the brain.

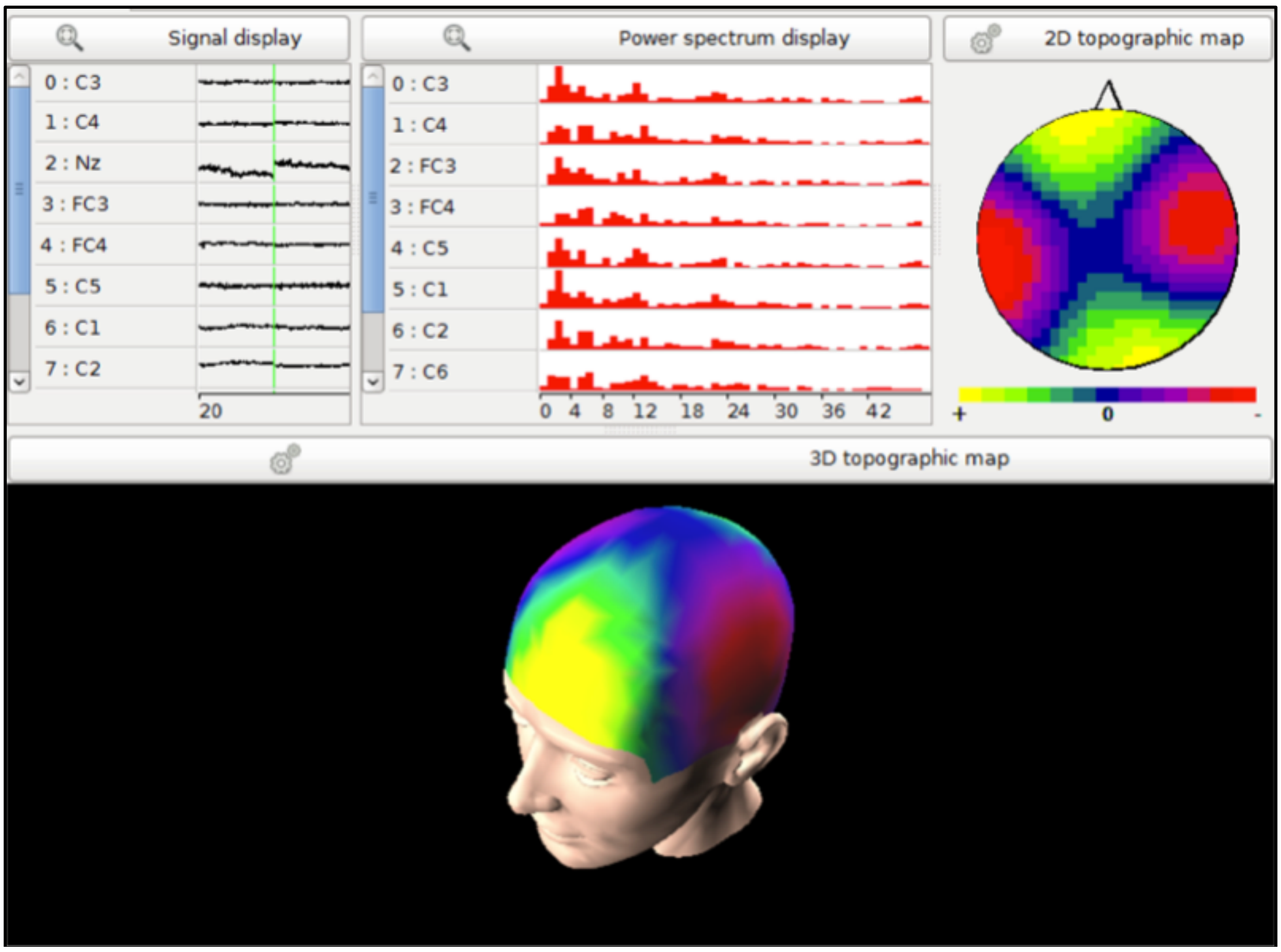

**Figure 4:** Snapshot of EEG activity mapped onto 2D and 3D topographical representations. (The electrodes’ labelling arrangement here is slightly different from the convention introduced in Figure 2.)

It is common knowledge that different regions of the brain have distinct roles; e.g., the visual cortex processes images, the auditory cortex processes sound, the motor cortex controls our limbs, and so on (Squire et al., 2008). Assorted regions of the brain cooperate to perform mental tasks. Moreover, spectral amplitudes of scalp EEG are constantly changing over the skull surface. This phenomenon is often compared to a weather system, where low and high pressures rotate masses of air over our planet. Therefore, states of mind cannot be accurately inferred simply by looking at a snapshot of the averaged EEG from the whole set of electrodes at once. Rather, one needs to look at the time-based interrelationships between spectral components of the EEG recorded simultaneously at different locations on the scalp.

## 3 An EEG-based logic of the mind

Petsche and Etlinger (1998) advocate that different bandwidths of the EEG spectrum are distinct windows to a dynamic landscape of electrical brain activity. It is suggested that mental states are correlated with interrelations between these windows. We follow this up by proposing that those interrelationships can be expressed by means of logical expressions. Hence, the notion of an EEG-based logic of the mind.

Consider a system, which extracts information from the EEG of a number of electrodes at given lapses of time; e.g., periods lasting for 500ms. For the sake of clarity, let us say that the system extracts beta and alpha rhythms from the EEG signals (Table 1). With this information, the system then tracks the behaviour of these rhythms over the scalp as time progresses. For this example, it tracks the electrodes' locations where beta and alpha rhythms displayed most power. This is shown in Figure 5: beta rhythms (denoted by the red electrodes) were detected prominently by electrode Fp2 at time $t^0$. Next, at time $t^1$ they were detected prominently by electrode Fz and then by electrode T4 at time $t^2$. And alpha rhythms (denoted by the blue electrodes) were detected prominently by electrodes T5, T3 and C3, respectively.

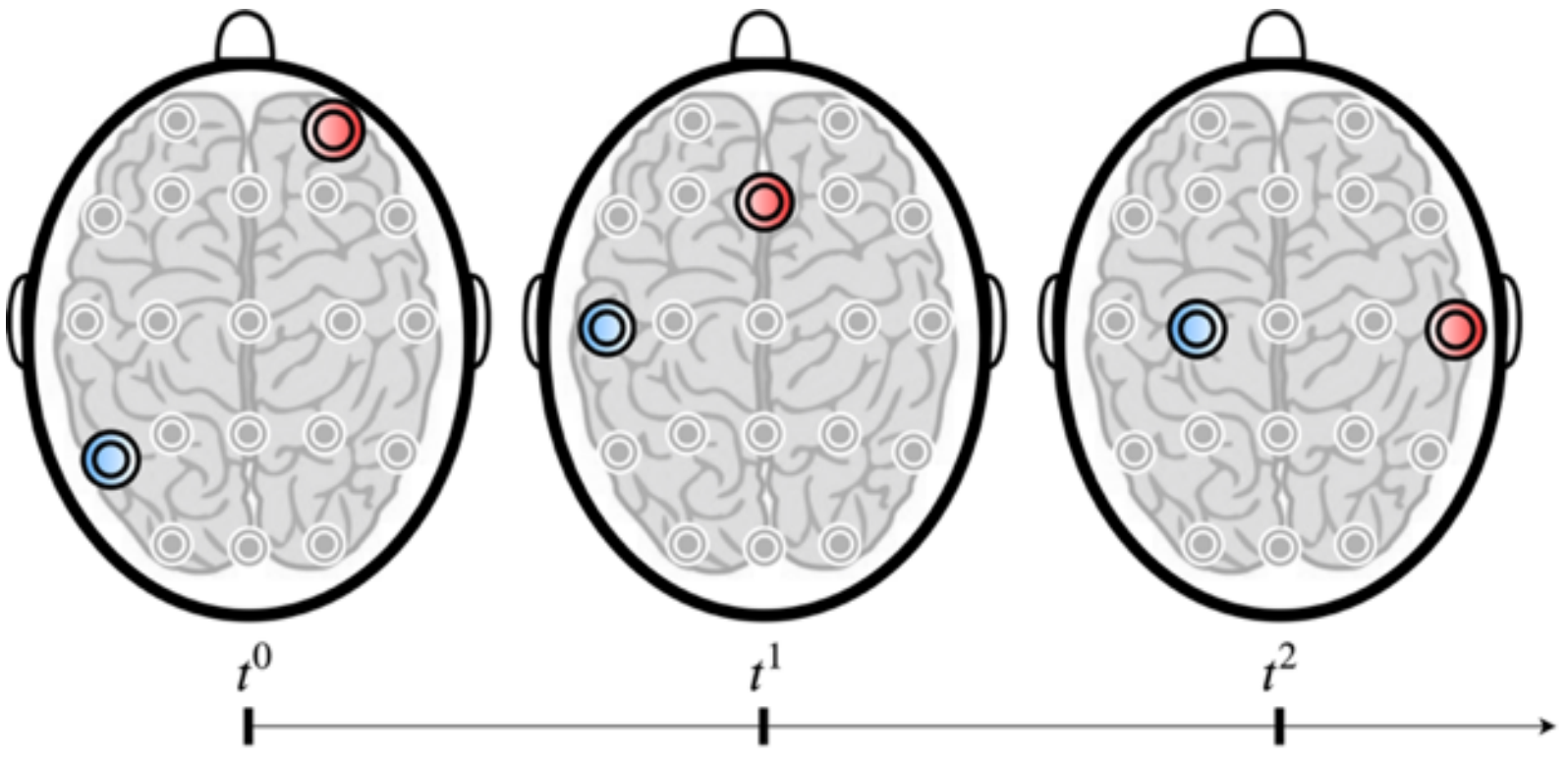

**Figure 5**: Tracking the trajectories of two EEG rhythms in time.

Next, at each time step $t^n$, the EEG information is encoded as a logic expression. The variables represent the activity of the respective electrodes.

In order to keep the example simple, let us consider only the following subset of electrodes: {Fp2, Fz, T3, C3, T4, T5}. In this case, the electrodes that detect the most prominent EEG signals are represented in the expression as 'True'. All the others are 'False'. Thus, the expressions for the time lapses in Figure 5 could be written follows[2]:

$t^0: \beta(\ \mathbf{Fp2} \wedge \neg \mathrm{Fz} \wedge \neg \mathrm{T3} \wedge \neg \mathrm{C3} \wedge \neg \mathrm{T4} \wedge \neg \mathrm{T5}) \wedge \alpha(\neg \mathrm{Fp2} \wedge \neg \mathrm{Fz} \wedge \neg \mathrm{T3} \wedge \neg \mathrm{C3} \wedge \neg \mathrm{T4} \wedge\ \mathbf{T5})$
$t^1: \beta(\neg \mathrm{Fp2} \wedge\ \mathbf{Fz} \wedge \neg \mathrm{T3} \wedge \neg \mathrm{C3} \wedge \neg \mathrm{T4} \wedge \neg \mathrm{T5}) \wedge \alpha(\neg \mathrm{Fp2} \wedge \neg \mathrm{Fz} \wedge\ \mathbf{T3} \wedge \neg \mathrm{C3} \wedge \neg \mathrm{T4} \wedge \neg \mathrm{T5})$
$t^2: \beta(\neg \mathrm{Fp2} \wedge \neg \mathrm{Fz} \wedge \neg \mathrm{T3} \wedge \neg \mathrm{C3} \wedge\ \mathbf{T4} \wedge \neg \mathrm{T5}) \wedge \alpha(\neg \mathrm{Fp2} \wedge \neg \mathrm{Fz} \wedge \neg \mathrm{T3} \wedge\ \mathbf{C3} \wedge \neg \mathrm{T4} \wedge \neg \mathrm{T5})$

The first term, inside the parenthesis on the left side of the conjunction operator, corresponds to the beta rhythms and the second term to the alpha ones.

[2] See (Smith, 2020) for an introduction to formal logic. The symbol ¬ is the negation operator (i.e., NOT), and ∧ stands for the logical conjunction operator (i.e., AND). Variables corresponding to a True electrode are in bold for clarity. These expressions could be stated in reduced form, but we leave them expanded here for the sake of intelligibility.

As a matter of fact, the expressions above could have been written with other logical operators. For instance, $t^0$ could have been written with the logical disjunction operator[3] between the beta and alpha terms:

$$t^k\colon \beta(\mathrm{Fp2} \wedge \neg\mathrm{Fz} \wedge \neg\mathrm{T3} \wedge \neg\mathrm{C3} \wedge \neg\mathrm{T4} \wedge \neg\mathrm{T5}) \vee \alpha(\neg\mathrm{Fp2} \wedge \neg\mathrm{Fz} \wedge \neg\mathrm{T3} \wedge \neg\mathrm{C3} \wedge \neg\mathrm{T4} \wedge \mathrm{T5})$$

The forms of those logical expressions depend on what they are meant to represent. They might, for example, depend on the correlated mental states which they are supposed to stand for. For instance, it is generally agreed that increased beta activity in the pre-frontal cortex corresponds to 'focused attention in problem-solving' (Ligeza et al., 2015). The pre-frontal cortex is covered by electrodes Fp1 and Fp2, and to some extent F7, F3, Fz, F4 and F8. Thus, the expressions for $t^0$ and $t^1$ above would correspond to (or 'satisfy' in logic parlance) this mental state, whereas the expression for $t^2$ would not.

The problem of establishing which values for the variables of a logical expression can render the whole expression True, or satisfiable, is known as the Boolean satisfiability problem. As an illustration, consider this simple expression, with variables A and B: $(\neg\mathrm{A} \wedge \mathrm{B})$. If A = False and B = True, then this expression is True. Hence, these values for A and B render the expression satisfiable. But if, say, B = False then the whole expression would be False.

The ability to represent EEG correlates in terms of logical expressions is useful for implementing BCI systems because they can be programmed to activate actions associated to specific expressions. As the EEG is acquired and analysed, a system would check when the information satisfies given logical expressions. Then, it would perform the respective actions for those expressions that return True. For example:

If at $t^n$, $\beta(a \wedge \neg(b \wedge c \wedge d \wedge e \wedge f)) \vee \alpha(\neg(a \wedge b \wedge c \wedge d \wedge e) \wedge f))$ = True
then { move robot arm to the left by 180° } ;

If at $t^n$, $\beta(b \wedge \neg(a \wedge c \wedge d \wedge e \wedge f) \wedge \alpha(\neg(a \wedge b \wedge c \wedge d \wedge f) \wedge e))$ = True
then { move robot arm to the left by 45° };

If at $t^n$, $\beta\big((a \vee d) \wedge \neg(b \wedge c \wedge e \wedge f)\big) \wedge \alpha(\neg(a \wedge b \wedge c) \wedge (d \vee e \vee f))$ = True
then { move robot arm to the right by 90° };

and so on.

Satisfiable complex Boolean expressions may return True for different combinations of logic values. And such expressions can have a great number of variables and a variety of satisfactory combinations.

In addition to building BCI systems, the proposed EEG-based logic of the mind is potentially useful for cataloguing mental states and EEG correlates. It is envisaged building systems programmed to automatically generate logical expressions while subjects are performing specific mental tasks. The hypothesis is that systems could evolve sophisticated time-based ontologies of mental activity in unprecedented ways.

[3] The symbol for the logical disjunction operator (i.e., OR) is $\vee$.

The caveat of developing a catalogue of mental states represented as logical expressions is that Boolean satisfiability problems are very demanding in terms of computation. These expressions could involve dozens of logical variables. Expressions with, say, 50 variables would require as many as $2^{50} = 1{,}125{,}899{,}906{,}842{,}624$ combinations to be checked. A modest personal computer is capable of performing 2 billion operations per second. Thus, such a computer would take 562,945 seconds to evaluate an expression. That is, almost an entire week. New quantum computing technology promises considerable speed-up for tasks such as these (Ball, 2014). Hence the rationale for researching the potential of quantum computing for BCI.

## 4 A brief introduction to quantum computing and logic operations

### 4.1 The basics of quantum computing

This section introduces the basics of quantum computing and logic operations. It focuses on the basics deemed necessary to contextualise and understand how to program quantum computers for solving logical expressions. For more detailed explanations of quantum computing, the reader is referred to (Johnston et al., 2019; Grumbling and Horowitz, 2019; Rieffel and Polak, 2011; Mermin, 2007; Bernhardt, 2019). A quantum computing primer is available in Chapter 1 in this book.

Classical computers manipulate information represented in terms of binary digits, each of which can value 1 or 0. They work with microprocessors made up of billions of tiny switches activated by electric signals. Values equal to 1 and 0 reflect the on and off states of the switches, respectively.

In contrast, a quantum computer deals with information in terms of quantum bits, or *qubits*. Qubits are subject to the laws of quantum mechanics because they operate at the subatomic level. At the subatomic level, a quantum object does not necessarily exist in a determined state. Its state is unknown until one observes it.

A qubit is a two-level quantum system where the two basis states are usually written as $|0\rangle$ and $|1\rangle$. A qubit can be in state $|0\rangle$, $|1\rangle$ or (unlike a classical bit) in a state of *superposition*, which is a linear combination of both. Superposition is the first fundamental property of quantum computing mentioned at the Introduction. To a greater extent, the art of programming a quantum computer involves manipulating qubits to perform operations while they are in such indeterminate state. This makes quantum computing fundamentally different from digital computing.

The difficulty with building quantum processors is that they need to be well isolated from the environment in order to keep the qubits coherent to perform computations. Quantum processors rely on qubits being kept in superposition for as long as possible. However, keeping a qubit in superposition is like balancing a tiny thin coin upright on a floating surface: any movement, even the tiniest vibration, will cause it to fall to head or tail. And this gets harder with groups of entangled qubits (e.g., when the state of a qubit depends on the state of another). This fall is referred to as decoherence.

Still, total isolation is impossible, because one needs to access the qubits in order to read information. The very act of reading qubits can change their state, because it is an intervention from the external world.

In order to picture a qubit, imagine a transparent sphere with opposite poles. From its centre, a vector whose length is equal to the radius of the sphere can point to anywhere on the surface. In quantum mechanics, this sphere is known as the Bloch sphere. And the vector is referred to as a state vector. The opposite poles of the sphere are denoted by $|0\rangle$ and $|1\rangle$, which is the notation used to represent quantum states (Figure 6).

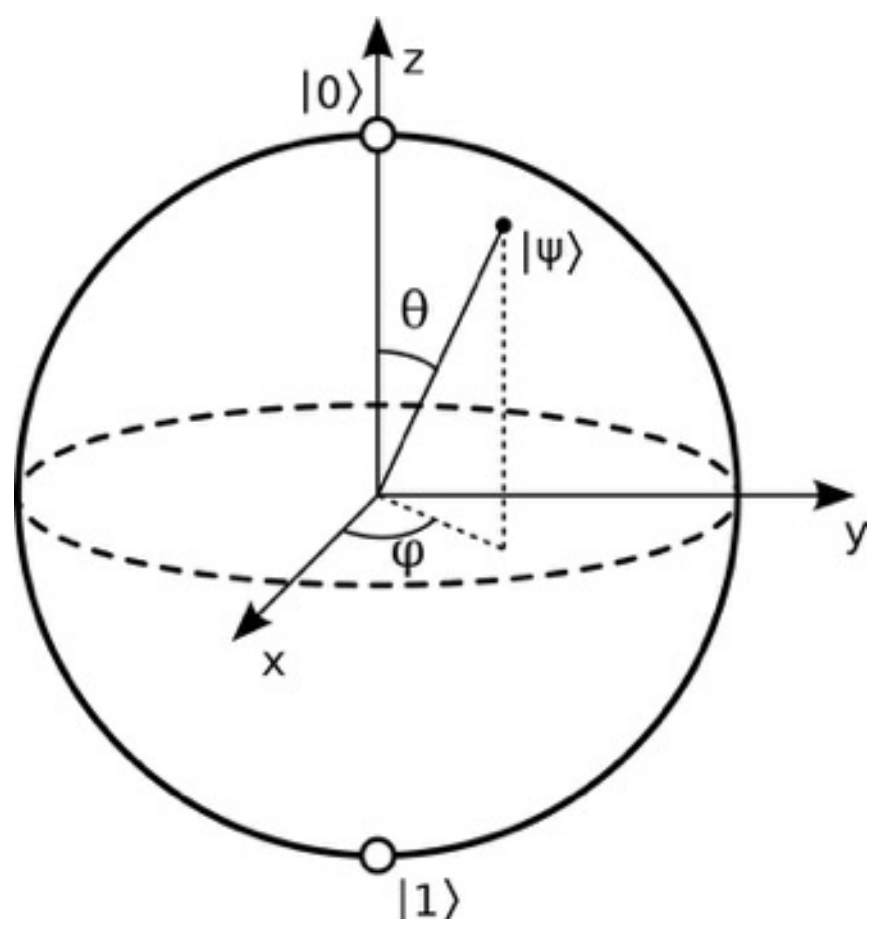

**Figure 6:** Bloch sphere. (Source: Smite-Meister, https://commons.wikimedia.org/w/index.php?curid=5829358)

A qubit's state vector can point at anywhere on the Bloch sphere's surface. Mathematically, this is described in terms of polar coordinates using two angles, $\theta$ and $\varphi$. The angle $\theta$ is the angle between the state vector and the z-axis (latitude) and the angle $\varphi$ describes the vector's position in relation to the x-axis (longitude).

When a qubit is in a state of superposition of $|0\rangle$ and $|1\rangle$, the state vector could be pointing anywhere between the two. However, we cannot really know where exactly a state vector is pointing to until we read the qubit. In quantum computing terminology, the act of reading a qubit is called 'measurement'. Measuring the qubit will make the vector point to one of the poles and return either 0 or 1 as a result.

The state vector of a qubit in superposition is described as a linear combination of two vectors, $|0\rangle$ and $|1\rangle$, as follows:

$$|\Psi\rangle = \alpha|0\rangle + \beta|1\rangle, \text{ where } |\alpha|^2 + |\beta|^2 = 1.$$

The state vector $|\Psi\rangle$ is a superposition of vectors $|0\rangle$ and $|1\rangle$ in a two-dimensional complex space, referred to as Hilbert space, with amplitudes $\alpha$ and $\beta$. In this case, the amplitudes are expressed in terms of Cartesian coordinates, which can be complex numbers. For example, consider the squared values of $\alpha$ and $\beta$ as probability values representing the likelihood of the measurement return 0 or 1. And let us assume the following:

$$|\Psi\rangle = \alpha|0\rangle + \beta|1\rangle, \text{ where } \alpha = \frac{1}{2} \text{ and } \beta = \frac{\sqrt{3}}{2}$$

In this case, $|\alpha|^2 = 0.25$ and $|\beta|^2$ = 0.75. Therefore, the measurement of the qubit has a 25% chance of returning 0 and a 75% chance of returning 1.

Quantum computers are programmed using sequences of commands, or quantum gates, that act on qubits. For instance, the 'not gate', performs a rotation of 180 degrees around the x-axis. Hence this gate is often called the X gate (Figure 7). A more generic rotational Rx($\vartheta$) gate is available for quantum programming, where the angle for the rotation is specified. Therefore, Rx(180) applied to $|0\rangle$ or $|1\rangle$ is equivalent to applying X to $|0\rangle$ or $|1\rangle$. In essence, all quantum gates perform qubit rotations, which change the amplitude distribution of the system.

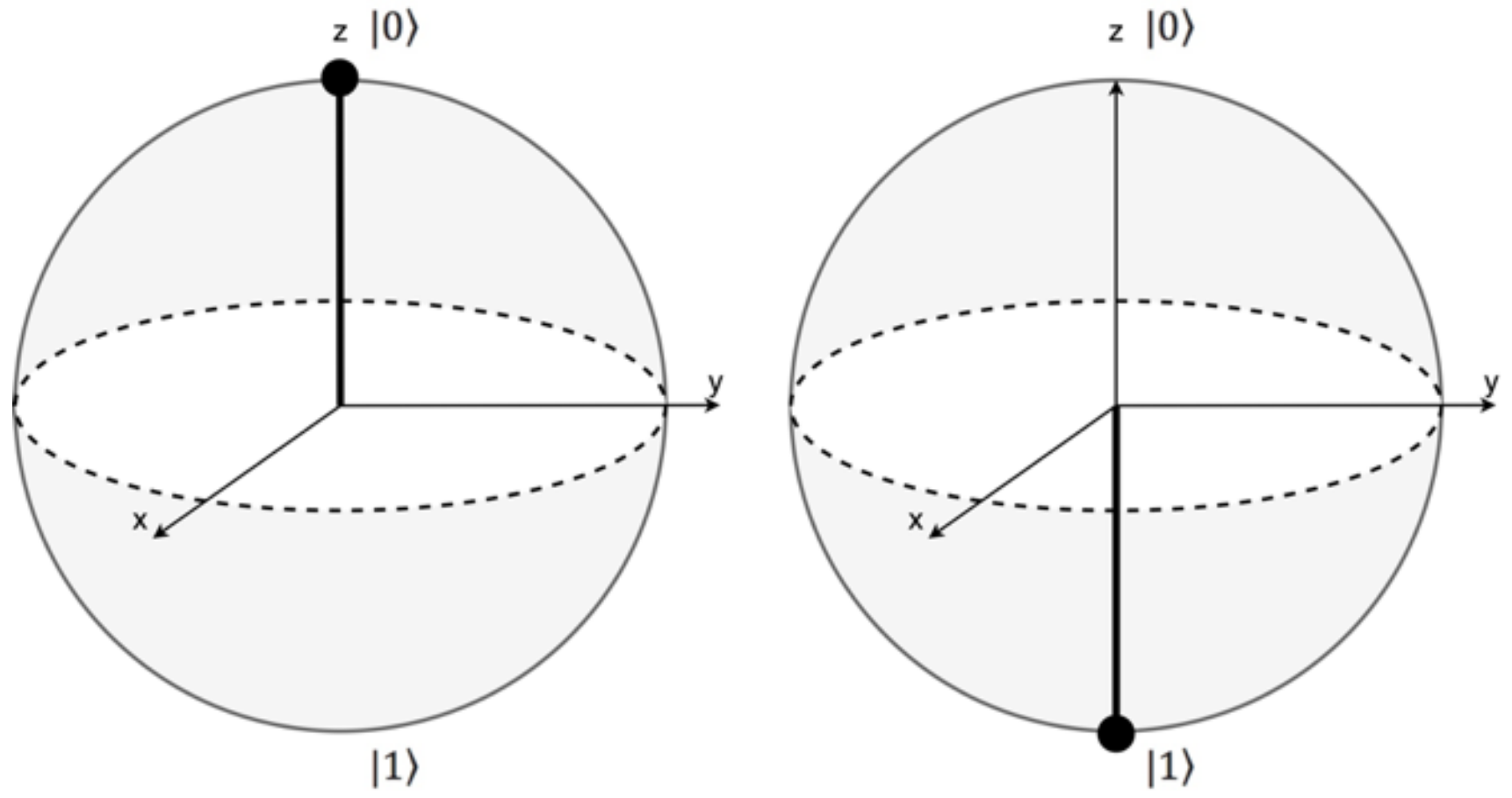

**Figure 7:** The X gate rotates a qubit's state vector (pointing upwards on the figure on the left side) by 180 degrees around the x-axis (pointing downwards on the figure on the right side).

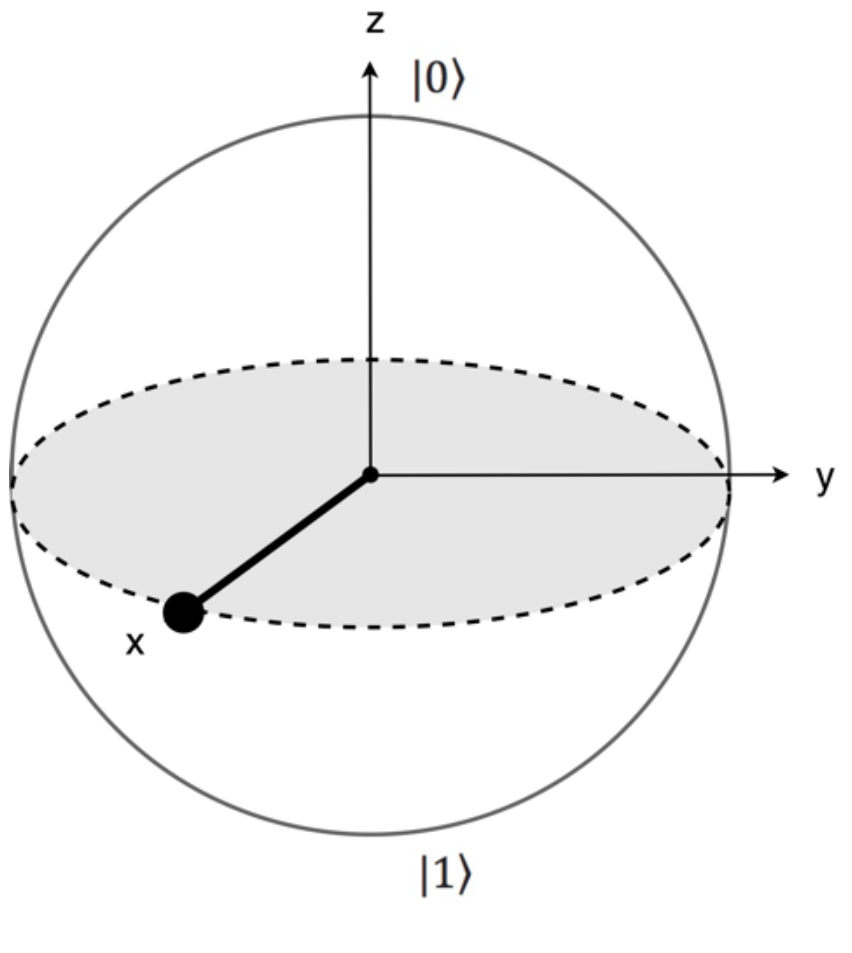

**Figure 8:** The Hadamard gate puts the qubit into a superposition state halfway two opposing poles.

An important gate is the Hadamard gate (referred to as the H gate). This gate puts the qubit into a superposition state consisting of an equal-weighted combination of two opposing states: $|\Psi\rangle = \alpha|0\rangle + \beta|1\rangle$ where $|\alpha|^2 = 0.5$ and $|\beta|^2 = 0.5$ (Figure 8). For other gates, please consult the references given earlier.

Qubits in a program typically start in ground state $|0\rangle$, and then a sequence of gates are applied. Then, the qubits are read and the results are stored in standard digital memory, which are accessible for further handling. A quantum program is often depicted as a circuit diagram of quantum gates, showing sequences of gate operations on the qubits (Figure 9).

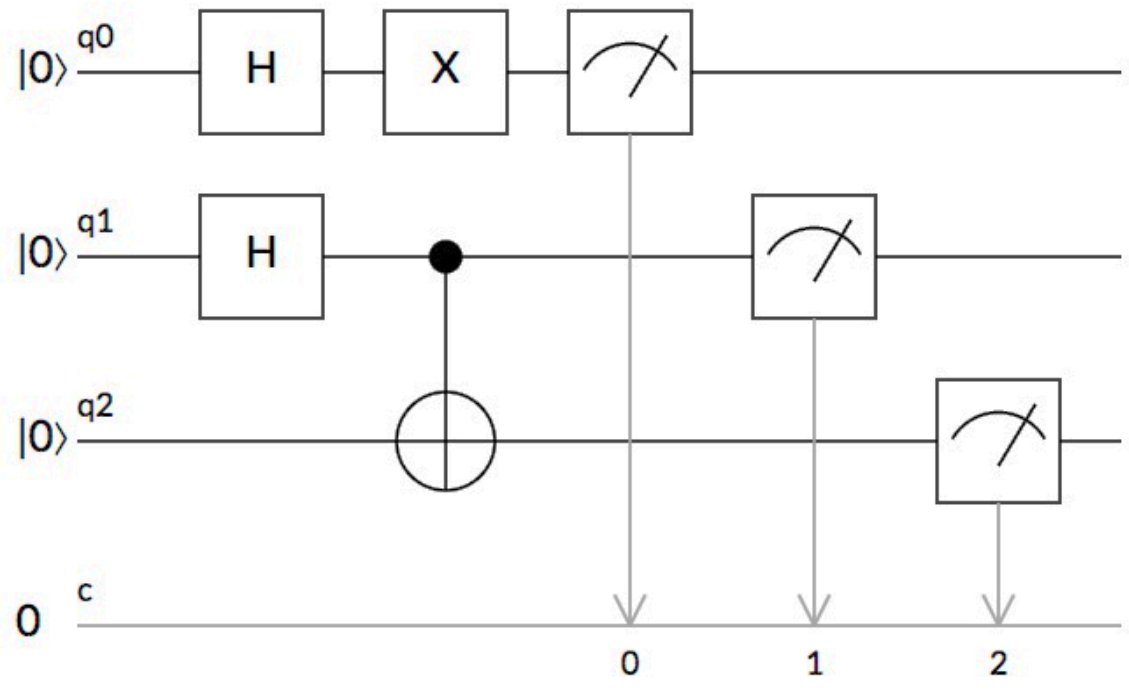

**Figure 9:** A quantum program depicted as a circuit of quantum gates. The squares with dials represent measurements, which are saved on classical registers represented at the bottom line.

The vertical z-axis of the Bloch sphere forms the so-called *standard computational basis*. The x-axis forms the *conjugate computational basis* and y-axis for the *circular computational basis*. As its name suggest, the standard basis is the most commonly used, and it is the one adopted for the work presented in this chapter. A detailed explanation of these bases and their significance to computation can be found in (Bernhardt, 2019). What is important to bear in mind here is that changing the basis on which a quantum state is expressed, corresponds to changing the measurement performed to read the outcomes of the computations. It is important to note that changing the basis to express a state does not change anything physical per se.

Quantum computation gets really interesting with gates that operate on multiple qubits, such as the conditional, or controlled, X gate; referred to as CX gate. The CX gate is an entangling gate. Depending on the input control state, it can put two qubits in *entanglement*, which is the second fundamental property of quantum computing mentioned at the Introduction. Entanglement establishes a curious correlation between qubits, which are impossible to establish between digital bits.

The CX gate applies an X gate on a qubit only if the state of another qubit is $|1\rangle$. Thus, the CX gate establishes a dependency of the state of one qubit with the value of another. The visual representation of the CX gate is shown in Figure 10. In fact, any quantum gate can be built in controlled form. And entanglement can take place between more than two qubits.

The Bloch sphere is useful for visualizing what happens with a single qubit, but it is not suitable for multiple qubits, in particular when they are entangled. Entangled qubits can no longer be thought of as independent units. They become one quantum entity described by a state vector of its own right. Hence, from now on we will have to use mathematics to represent quantum systems.

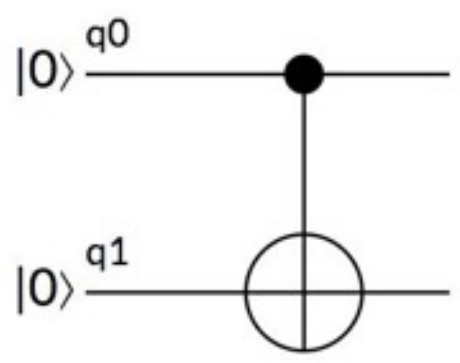

**Figure 10:** The CX gate creates a dependency of the state of one qubit with the state of another. In this case, q1 will be flipped only if q0 is $|1\rangle$.

The notation used above to represent quantum states ($|\Psi\rangle, |0\rangle, |1\rangle$), is called Dirac notation, which provides an abbreviated way to represent vectors. For instance, $|0\rangle$ and $|1\rangle$ represent the following vectors, respectively:

$$|0\rangle = \begin{bmatrix} 1 \\ 0 \end{bmatrix} \quad \text{and} \quad |1\rangle = \begin{bmatrix} 0 \\ 1 \end{bmatrix}$$

And quantum gates are represented as matrices. For instance, the X gate is represented as:

$$X = \begin{bmatrix} 0 & 1 \\ 1 & 0 \end{bmatrix}$$

Mathematically, quantum gate operations are expressed as matrix operations. Thus, the application of an X gate to $|0\rangle$ is the multiplication of a matrix (gate) by a vector (qubit state), which looks like this:

$$X(|0\rangle) = \begin{bmatrix} 0 & 1 \\ 1 & 0 \end{bmatrix} \times \begin{bmatrix} 1 \\ 0 \end{bmatrix} = \begin{bmatrix} 0 \\ 1 \end{bmatrix} = |1\rangle$$

Quantum processing with multiple qubits is represented by means of tensor vectors. A tensor vector is the result of the tensor product, represented by the symbol $\otimes$, of 2 or more vectors. A system of two qubits looks like this $|0\rangle \otimes |0\rangle$, but it is normally abbreviated to $|00\rangle$. However, it is useful to look at the expanded form of the tensor product to trace how it works:

$$|00\rangle = |0\rangle \otimes |0\rangle = \begin{bmatrix} 1 \\ 0 \end{bmatrix} \otimes \begin{bmatrix} 1 \\ 0 \end{bmatrix} = \begin{bmatrix} 1 \times 1 \\ 1 \times 0 \\ 0 \times 1 \\ 0 \times 0 \end{bmatrix} = \begin{bmatrix} 1 \\ 0 \\ 0 \\ 0 \end{bmatrix}$$

The CX gate, for instance, is defined by the matrix:

$$CX = \begin{bmatrix} 1 & 0 & 0 & 0 \\ 0 & 1 & 0 & 0 \\ 0 & 0 & 0 & 1 \\ 0 & 0 & 1 & 0 \end{bmatrix}$$

Thus, application of CX to $|10\rangle$ is represented as:

$$CX(|10\rangle) = \begin{bmatrix} 1 & 0 & 0 & 0 \\ 0 & 1 & 0 & 0 \\ 0 & 0 & 0 & 1 \\ 0 & 0 & 1 & 0 \end{bmatrix} \times \begin{bmatrix} 0 \\ 0 \\ 1 \\ 0 \end{bmatrix} = \begin{bmatrix} 0 \\ 0 \\ 0 \\ 1 \end{bmatrix}$$

The resulting vector is then abbreviated to $|11\rangle$ as show below:

$$\begin{bmatrix} 0 \\ 0 \\ 0 \\ 1 \end{bmatrix} = \begin{bmatrix} 0 \\ 1 \end{bmatrix} \otimes \begin{bmatrix} 0 \\ 1 \end{bmatrix} = |1\rangle \otimes |1\rangle = |11\rangle$$

Table 2 shows the resulting quantum states of CX gate operations, where the second qubit flips only if the first qubit is $|1\rangle$. Note that in quantum computing, qubit strings are usually enumerated from the right end of the string to the left: e.g., $|q_2\rangle \otimes |q_1\rangle \otimes |q_0\rangle$. This is the norm adopted in this chapter from now on. Thus, the 'first qubit' is the rightmost one.

| Input | Result |
|---|---|
| $\|00\rangle$ | $\|00\rangle$ |
| $\|01\rangle$ | $\|11\rangle$ |
| $\|10\rangle$ | $\|10\rangle$ |
| $\|11\rangle$ | $\|01\rangle$ |

**Table 2:** CX gate table, where q1 is flipped only if q0 is $|1\rangle$.

Another useful controlled gate is the multiple-controlled form of the X gate, also known as the Toffoli gate (Aharonov, 2003). An example of a 3-qubit Toffoli gate (also known as CCX or CCNOT) in shown in Figure 11.

Table 3 shows resulting quantum states of the 3-qubit Toffoli gate portrayed in Figure 11.

| Input | Result |
|---|---|
| $\|000\rangle$ | $\|000\rangle$ |
| $\|001\rangle$ | $\|001\rangle$ |
| $\|010\rangle$ | $\|010\rangle$ |
| $\|011\rangle$ | $\|111\rangle$ |
| $\|100\rangle$ | $\|100\rangle$ |
| $\|101\rangle$ | $\|101\rangle$ |
| $\|110\rangle$ | $\|110\rangle$ |
| $\|111\rangle$ | $\|011\rangle$ |

**Table 3:** The 3-qubit Toffoli gate table where q2 flips only if both, q0 and q1 are $|1\rangle$.

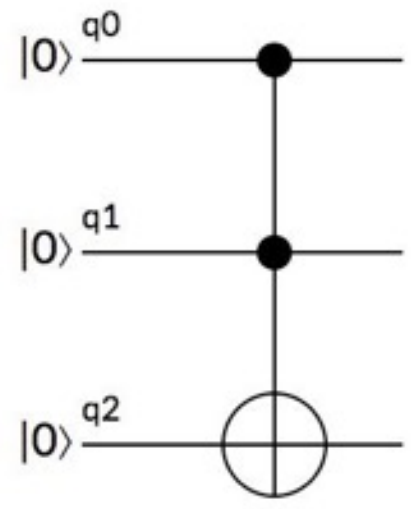

**Figure 11:** The 3-qubit Toffoli gate creates a dependency of the state of one qubit with the state of two qubits. In this case, q2 flips only if both, q0 and q1 are |1⟩.

## 4.2 Quantum logic operators

In digital logic, one can build any logic operator and entire Boolean expressions with just one basic NAND operator (Akerkar and Akerkar, 2004). Likewise, quantum logic operators and expressions can be built using only the X gate and its controlled forms. A few examples of such quantum logic operators are shown in Figure 12.

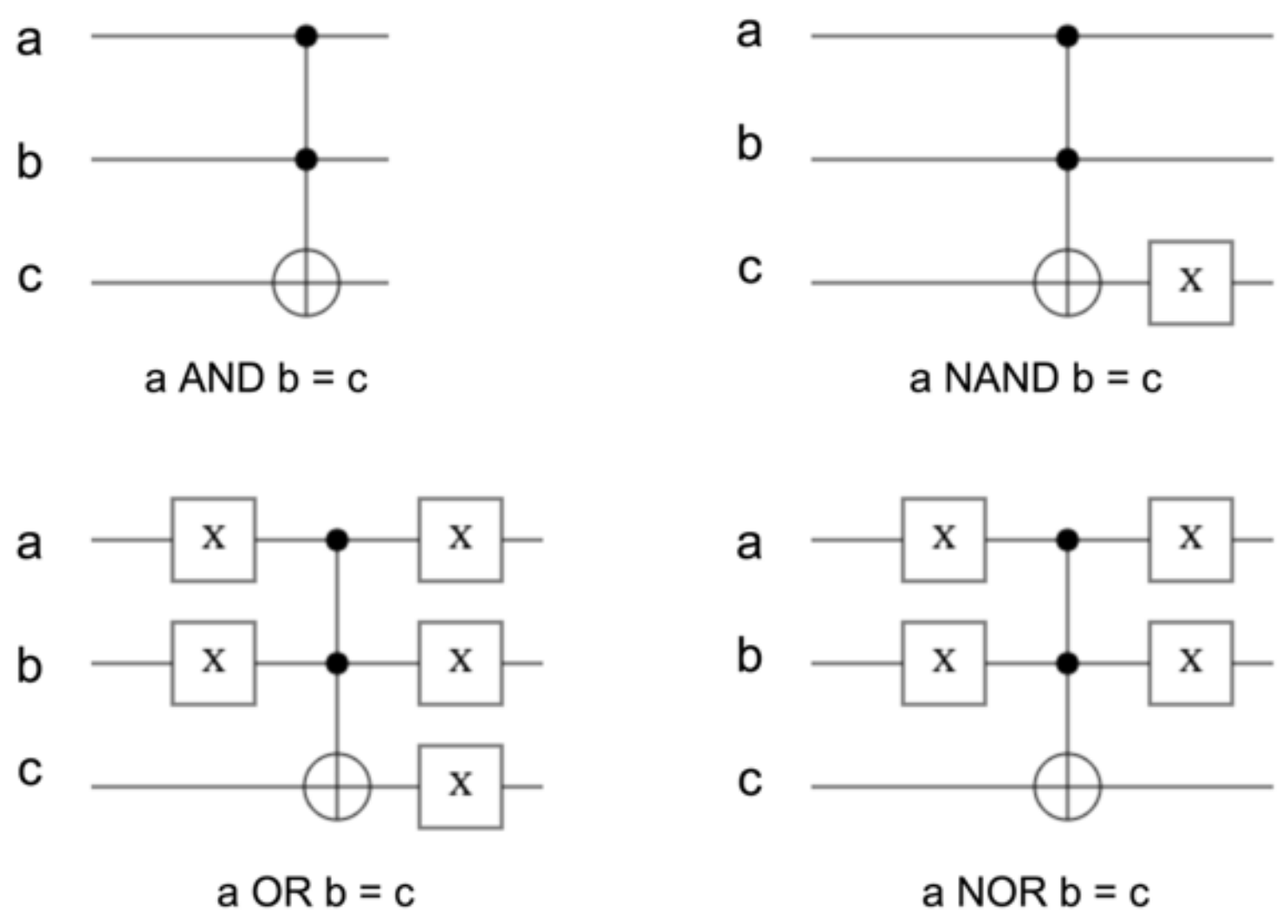

**Figure 12:** Quantum logic operators built with X gates. It is assumed that qubit c is initialized to |0⟩.

In addition to the X gate, another useful basic gate with which we can build Boolean expressions is the Rz($\vartheta$) gate. This gate rotates the state vector of a qubit around the z-axis of the Bloch sphere by a given angle $\vartheta$. In cases where the angle $\vartheta$ is equal to 180° then the Rz($\vartheta$) gate is often called as the Z gate (Figure 13).

Rotations around the z-axis represent changes in the phase of a qubit. For a qubit in uniform superposition with an equal probability to be measured |0⟩ or |1⟩, the state vector will point to the equator line. The Z gate thus reverses the phase of the qubit, while maintaining its superposition.

Boolean gates built using controlled Z gates encode the results of operations in the phases of the qubits. This allows for the representation of multiple outcomes in superposition, which is something that cannot be done using digital bits. Hence the beauty and processing power of quantum computing.

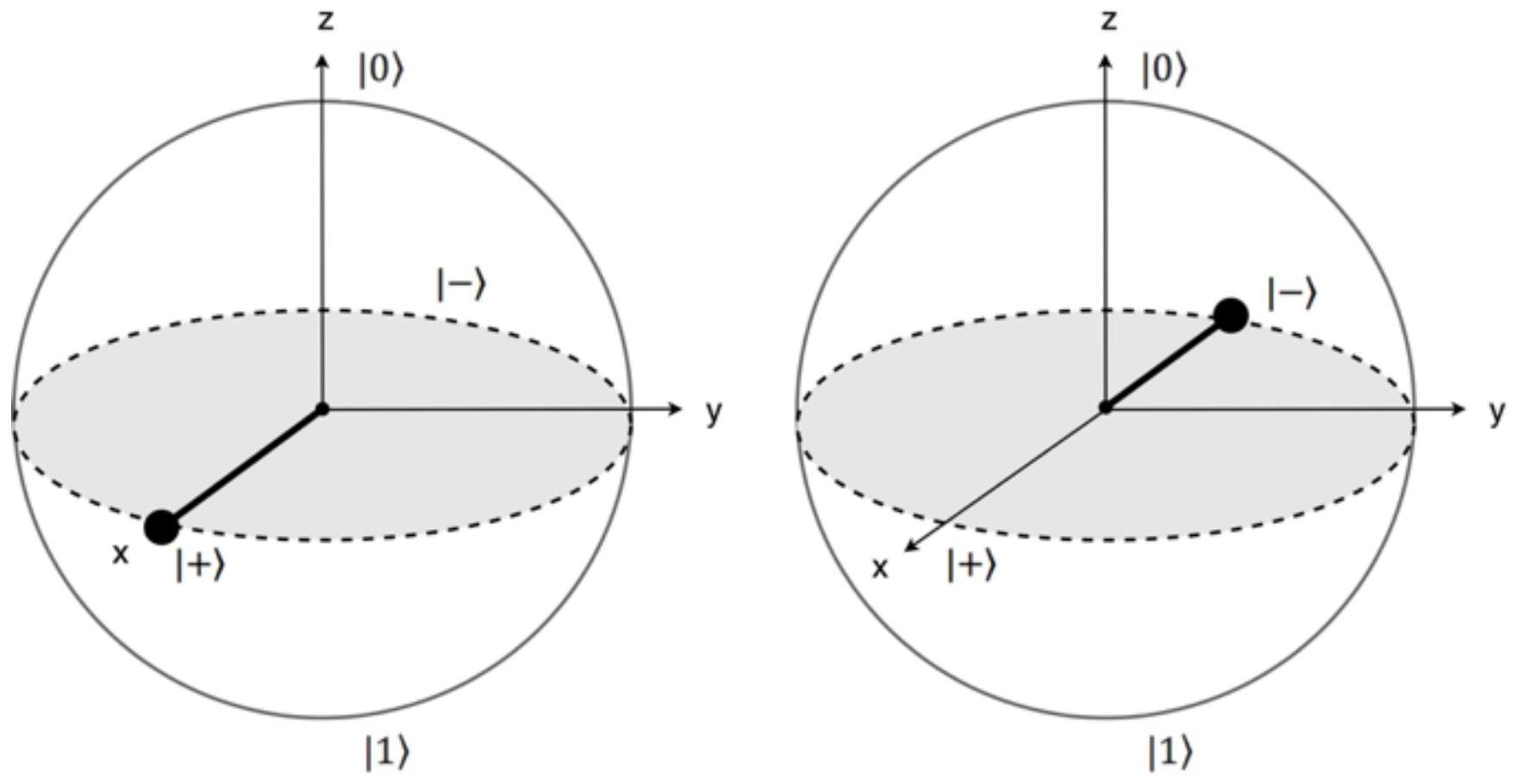

**Figure 13:** The Z gate rotates a qubit's state vector by 180 degrees around the z-axis. The opposite ends for this state vector are notated as $|+\rangle$ and $|-\rangle$.

A controlled Z gate acts only when both, the control qubit(s) and the target qubit are pointing to $|1\rangle$. Therefore, the logic circuit will change the phases only of those qubits that satisfy the operations they represent. These changes are referred to as 'spin-marking' the possible outcomes. Examples of logic operators built with controlled Z gates are shown in Figure 14.

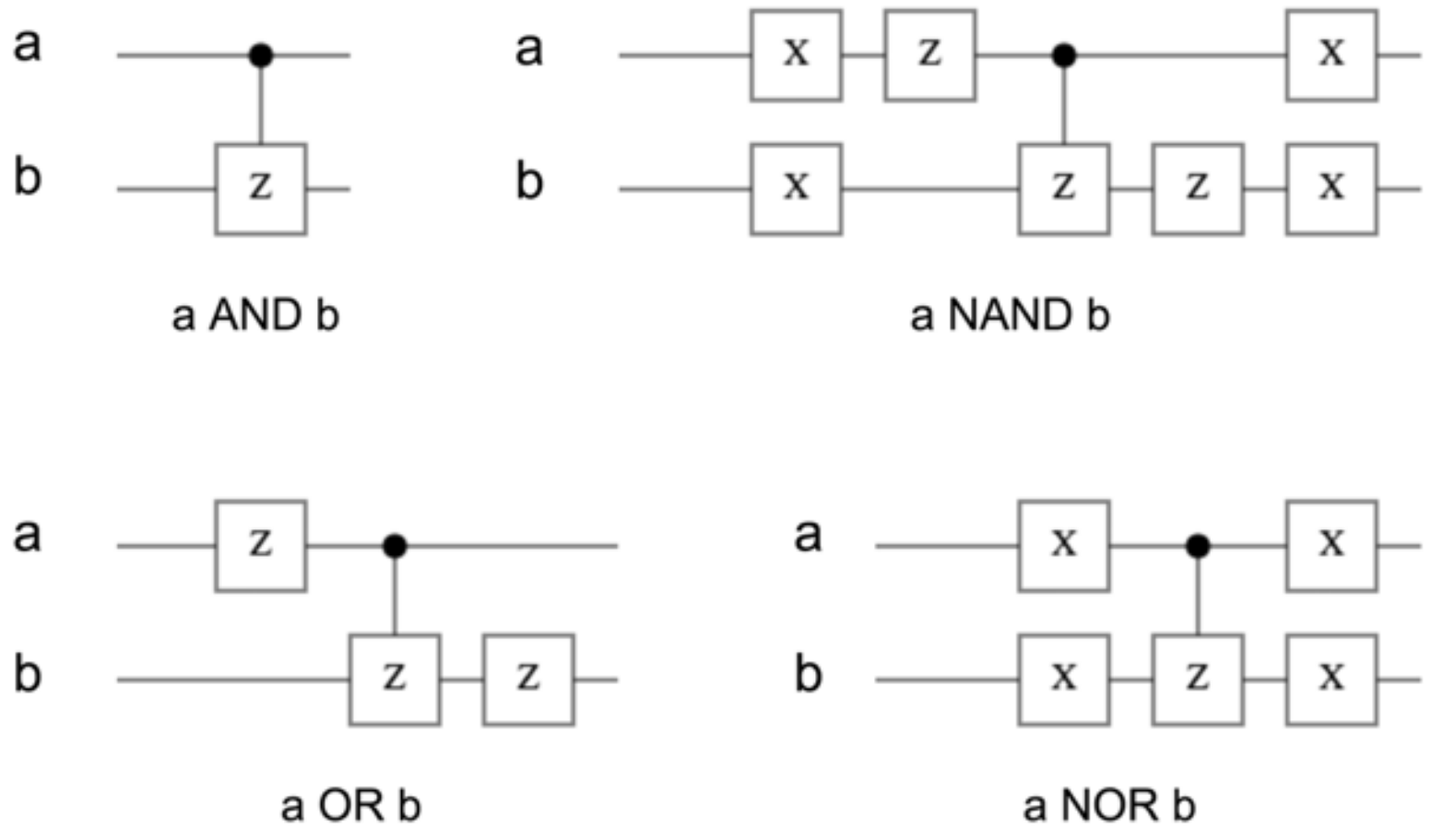

**Figure 14:** Quantum phase logic operators using Z gates.

Notice that one cannot simply link sequences of phase logic statements in a circuit. Their 'inputs' and 'outputs' are incompatible because they operate on different computational basis.

Recall that measurements force the qubits of a quantum system to settle to $|0\rangle$ or $|1\rangle$, according to the amplitudes of a wavefunction, which describes the system. The amplitudes represent the probability of the qubits returning 0 or 1. But rotations on the z-axis do not affect the amplitudes of the wavefunction. If one measures a phase logic operator on the standard basis, phase information is lost. Hence, in order to be extracted, phase information needs to be converted into amplitude information. This is done by means of a technique known as amplitude amplification.

An amplitude amplifier is a device that increases the probability of revealing which qubits have been spin-marked by the Z gates in a logical operation. The combination of spin-marking and amplitude amplification illustrates the third fundamental property of quantum computing that makes it different from classical computing: *interference*. Figure 15 shows how the circuit for amplitude amplification looks like. In this case, the circuit is for three qubits; if more are needed, then identical gate sequences are added for each additional qubit.

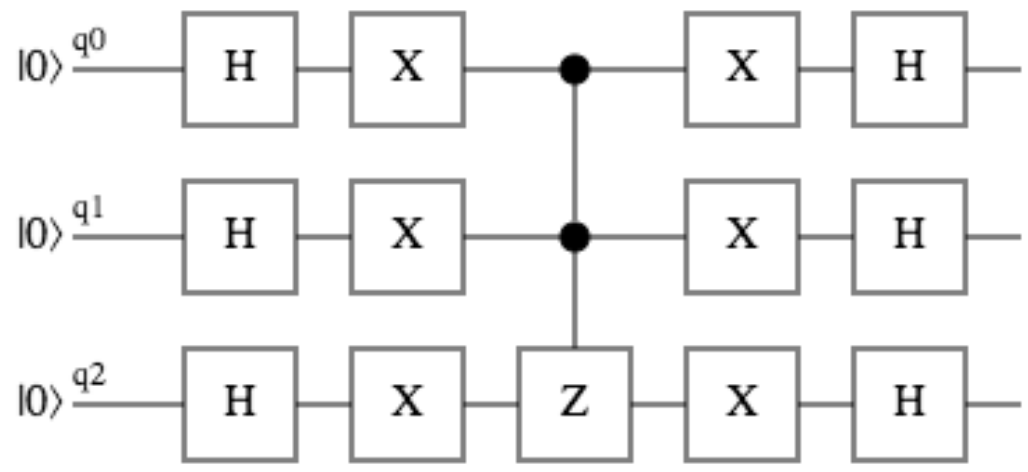

**Figure 15:** The amplitude amplifier circuit[4].

## 5 An example of a system: listening to the quantum logic of the mind

This section demonstrates how the concepts introduced above fit together. It presents a BCI system that reads the EEG of a user at given lapses of time and builds logic expressions encoding the activity of the EEG. Then, the system generates an equivalent circuit and submits it to a quantum computer to check the satisfiability of the expression. The result is then converted into sound, which is produced while the system deals with the EEG of the next lapse. And the cycle continues until the user halts it (Figure 16).

Note that instead of associating logical expressions with actions to be performed through the BCI, this section introduces a slightly different application for the logic of the mind: to build BCI-based musical instruments.

### 5.1 Building logic expressions from EEG information

Let us begin by defining a format for representing EEG information as logic expressions with only three clauses and three logical variables (A, B, and C) as follows:

$$(Clause_1) \wedge (Clause_2) \wedge (Clause_3)$$

[4] In the case of controlled Z gates, the black dot and the boxed circuit representation have the same effect; sometimes only a vertical line of black dots are used.

Each clause has two terms of the form $(Term_1 \vee Term_2)$. Below are examples of logic expressions in the proposed format:

$$(\neg A \vee C) \wedge (\neg B \vee \neg C) \wedge (A \vee C)$$

$$(A \vee B) \wedge (\neg B \vee C) \wedge (A \vee \neg C)$$

$$(A \vee B) \wedge (\neg B \vee \neg C) \wedge (A \vee C)$$

Next, consider that the clauses represent EEG information captured by electrodes positioned at specific places on the head (Figure 3), as shown in Table 3 and illustrated in Figure 17.

| | **A** | **B** | **C** |
|---|---|---|---|
| $\boldsymbol{Clause_1}$ | Fp1 | T3 | O1 |
| $\boldsymbol{Clause_2}$ | Fz | Cz | Oz |
| $\boldsymbol{Clause_3}$ | Fp2 | T4 | O2 |

**Table 3:** Specific electrodes are allocated to distinct A, B and C variables for each clause.

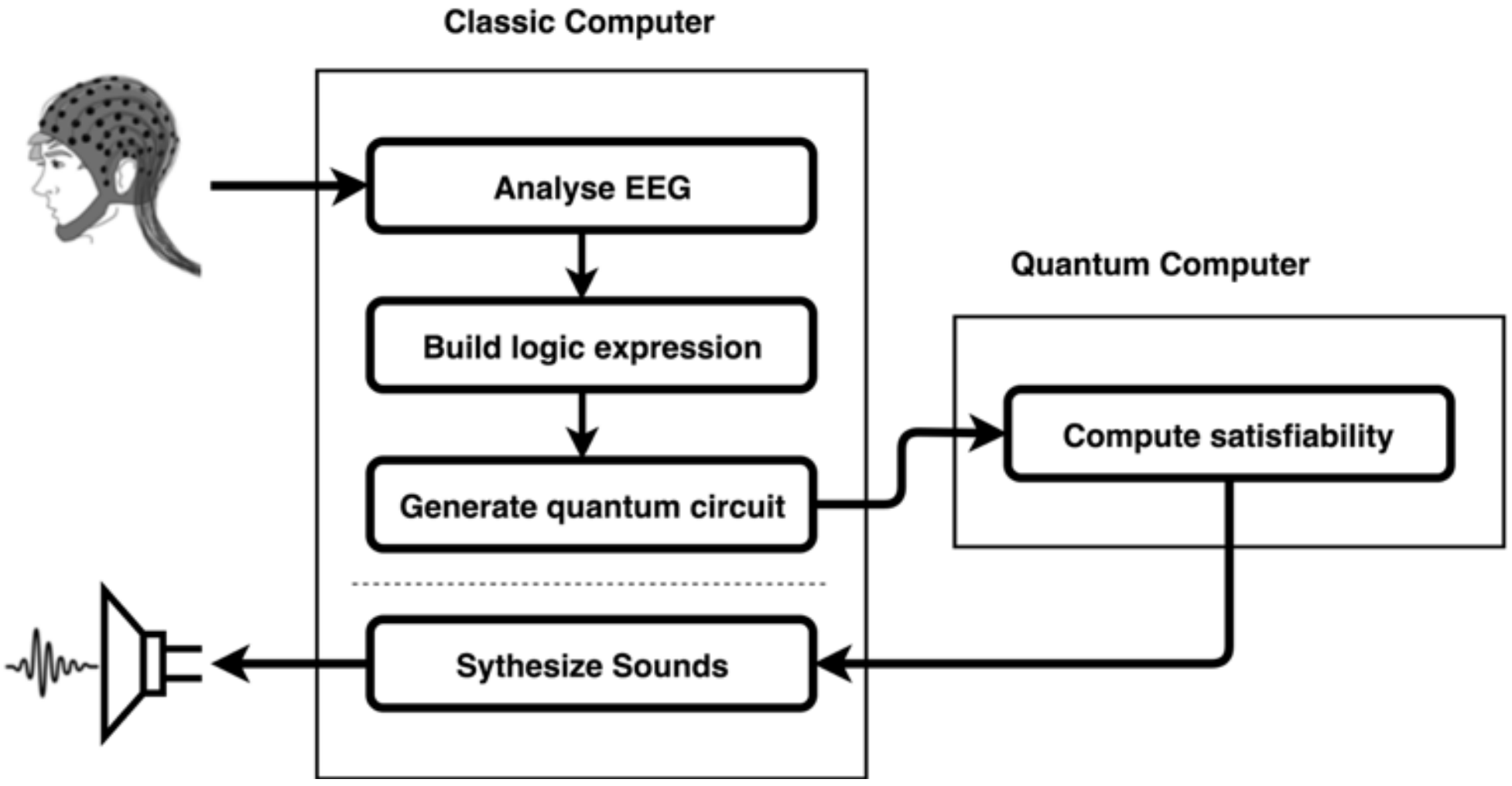

**Figure 16:** The example system architecture.

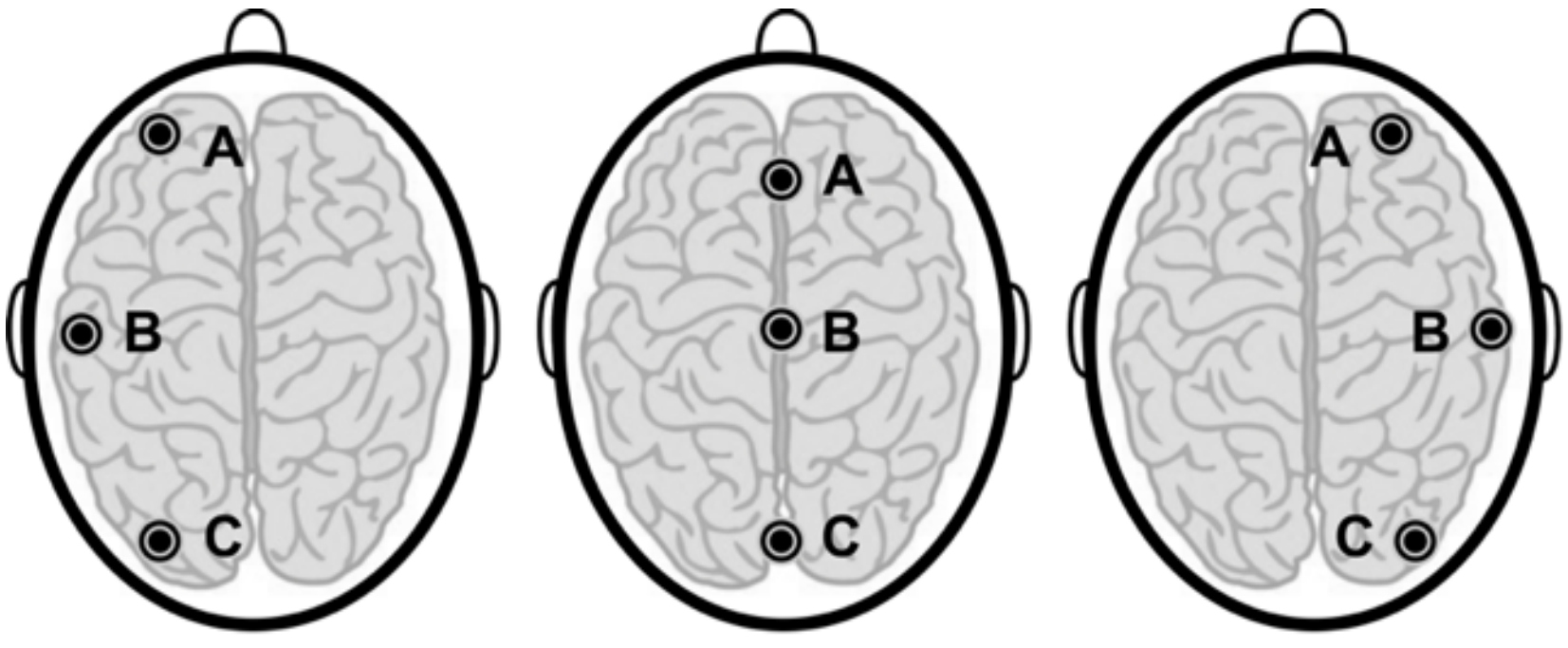

**Figure 17:** Electrodes and respective logic variables allocations.

Let us suppose that we are dealing with a BCI system that banks on EEG beta rhythms in order to activate commands to control an hypothetical system.

The terms of the clauses are stipulated as follows: for each clause, the system selects the two electrodes with the two highest EEG amplitudes. As an example, let us pretend that at a certain moment in time the electrodes Fp1 and O1 registered the highest amplitudes. Then, the system analyses the spectrum of the EEG from these two electrodes and extracts the most prominent frequency component in the spectrum, for each electrode. Let us say that the strongest component for Fp1 was 33.18 Hz and for O1 was 23.61 Hz. These correspond to terms A and C of $Clause_\mathbf{1}$; that is, A = 33.18 Hz and C = 23.61 Hz.

Next, the system checks if the frequencies conveyed by the respective terms are beta rhythms. If a frequency is equal to, or higher than, 15 Hz, then the respective term is True. Otherwise, it is False. In this case A = True and C = True. Therefore, these two terms form the clause $(\mathrm{A} \vee \mathrm{C})$. Had Fp1 been equal to, say, 10.36 Hz, then this term would have been $(\neg\mathrm{A} \vee \mathrm{C})$ instead.

In order to continue developing our example, let us suppose that the EEG analysis resulted in the following expression :

$$(\mathrm{A} \vee \mathrm{B}) \wedge (\neg\mathrm{B} \vee \neg\mathrm{C}) \wedge (\mathrm{A} \vee \mathrm{C})$$

At this stage the system verifies if this expression is satisfiable. In other words, it checks if those EEG values would render this expression True. Let us examine how the system generates a quantum circuit to check the satisfiability of our expression.

## 5.2 Generating quantum circuits for logic satisfiability

The system uses X and Z gates, and their multiple-controlled forms, to build logical expressions with the format introduced above. They are formed by three clauses linked by conjunction operators. And each of the three clauses are formed by terms linked by disjunction operators. The three disjunction clauses are built with X gates. And then, they are linked with conjunction operators implemented with a controlled Z gate.

The circuit requires only six qubits: three of them (q0, q1, and q2) represent the logical variables A, B, and C. The other three (q3, q4 and q5) are ancillary qubits, which are used to represent the results of the disjunction clauses. The full circuit is shown in Figure 18 and the respective Quil[5] code in Code 1.

To begin with, all qubits are initialized to $|0\rangle$ and the ones representing the logical variables are put in uniform superposition with the H gate.

The thee disjunction clauses are specified, one after the other, as shown in dashed boxes at the top of Figure 18. The the results from each clause are held in ancillary qubits: N1 in q3, N2 in q4, and N3 in q5. Then, the tripartite conjunction operation is implemented with a 3-qubit controlled Z gate applied to the ancillary qubits: they will be spin-marked by the phase-logic AND operation (Figure 14).

```
DECLARE ro BIT[3]
H 0
H 1
H 2
X 0
X 1
CCNOT 0 1 3
X 0
X 1
X 3
X 1
X 2
X 1
X 2
CCNOT 1 2 4
X 1
X 2
X 4
X 1
X 2
X 0
X 2
CCNOT 0 2 5
```

```
X 0
X 2
X 5
CONTROLLED CONTROLLED Z 3 4 5
X 0
X 2
X 5
CCNOT 0 2 5
X 0
X 2
X 1
X 2
X 1
X 2
X 4
CCNOT 1 2 4
X 1
X 2
X 1
X 2
X 0
X 1
X 3
```

```
CCNOT 0 1 3
X 0
X 1
H 0
H 1
H 2
X 0
X 1
X 2
CONTROLLED CONTROLLED Z 0 1 2
X 0
X 1
X 2
H 0
H 1
H 2
MEASURE 0 ro[0]
MEASURE 1 ro[1]
MEASURE 2 ro[2]
```

**Code 1:** Generated Quil code for the circuit in Figure 18. CCNOT stands for CCX, or Toffoli gate.

Next, the three disjunction clauses need to be uncomputed in order to return the ancillary qubits to their initial states. Uncomputation is achieved by running the respective logic sub-circuits back to front; the 'Inverted clauses' in the middle section of Figure 18. Then, amplitude amplification is applied to the logic variables to reveal the spin-marked qubits in terms of amplitudes. Finally, the qubits representing the logic values are measured to yield the result.

The ancillary qubits are not part of the result. Therefore, they are not measured in the end. However, they can interfere with the measurements because they are entangled with the qubits that represent the logical variables. Hence, the ancillary qubits must be uncomputed to make them return to their unentangled state[6].

[5] Quil is a quantum instruction language developed by Rigetti.
[6] Refer to (Johnston et al., 2019) for a didactic discussion on uncomputing ancillary qubits and implications to measurement.

In this example, the circuit spin-marks the quantum states $|001\rangle$, $|011\rangle$ and $|101\rangle$. Thus, each time the algorithm runs, it will output either 001, 011 and 101 with much higher probability than any of the other possible values. The plot in Figure 19 shows the times that each of the possible outcomes were observed after running and measuring the circuit for 5,000 times, or 5,000 'shots' in quantum computing terminology, on a Rigetti's Quantum Virtual Machine (QVM)[7]. See Appendix I for more examples.

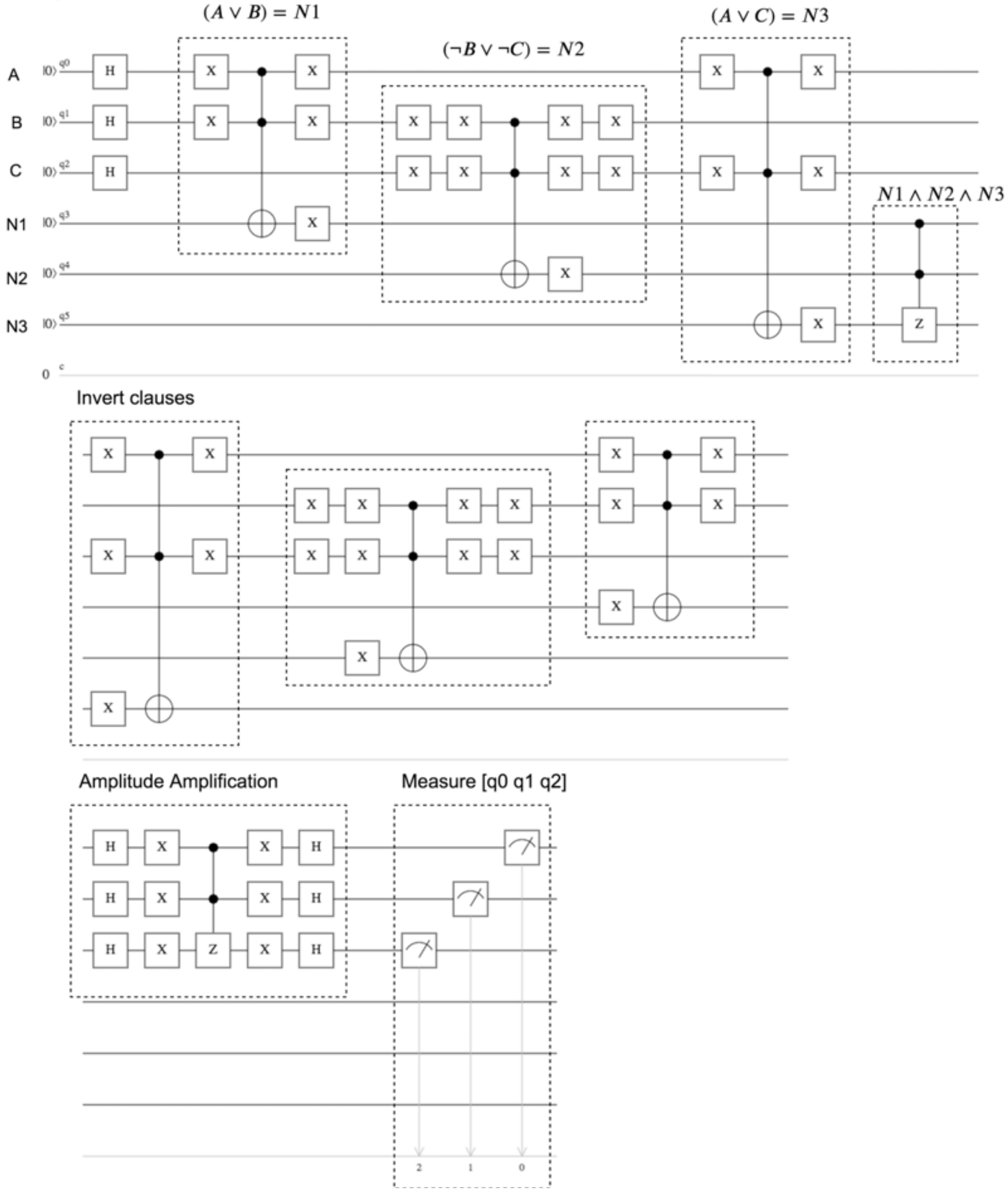

**Figure 18:** Quantum circuit for the expression $(A \vee B) \wedge (\neg B \vee \neg C) \wedge (A \vee C)$.

7 The Rigetti Quantum Virtual Machine is an implementation of the Quantum Abstract Machine described in (Smith et al., 2017). Noise simulation models to account for the effect of quantum hardware decoherence were not used in the examples discussed in this chapter.

What does this result mean? Bearing in mind that in quantum computing qubit strings are enumerated from the right end of the string to the left (i.e., $|\mathrm{CBA}\rangle$), the expression $(\mathrm{A} \vee \mathrm{B}) \wedge (\neg\mathrm{B} \vee \neg\mathrm{C}) \wedge (\mathrm{A} \vee \mathrm{C})$ is satisfied when:

Case 1, $|001\rangle$: A = True, B = False and C = False
Case 2, $|011\rangle$: A = True, B = True and C = False
Case 3, $|101\rangle$: A = True, B = False and C = True

In other words, should there be a BCI command associated with this expression, then it would be triggered when significant beta rhythms are detected: (a) by electrodes Fp1 and Fp2 (Case 1); or (b) by electrodes Fp1, Fp2 and T3 (Case 2); or (c) by electrodes Fp1, Fp2 and O2 (Case 3).

In terms of mental states, this expression encodes a state of focused attention. It is well known that beta rhythms in the frontal regions of the brain are associated with this mental state (Berta et al., 2013). A BCI system could be programmed to execute a certain task when the user is focusing attention on something.

Whereas an in-depth account of mental states associated to EEG is far beyond the scope of this chapter, the example above offers a glimpse into the potential for neurotechnology of the proposed logical of the mind with quantum computing.

The next section describes our method to render into sound the outcomes from running a quantum circuit to check the satisfiability of a logic expression, over a number of shots.

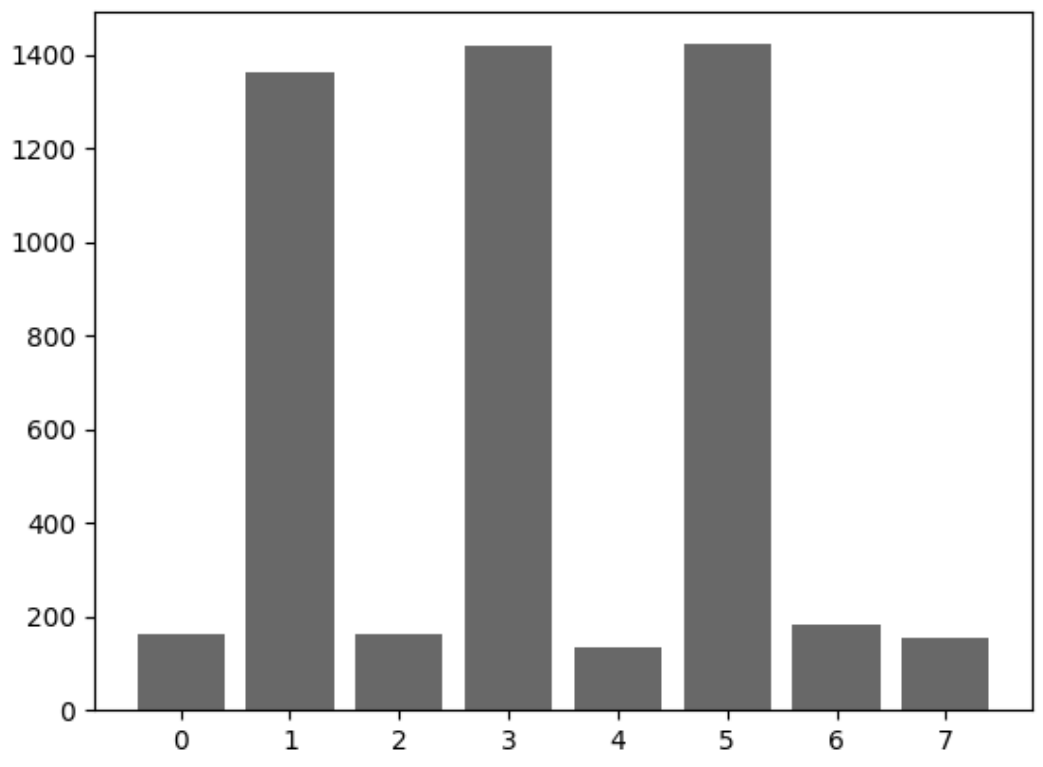

**Figure 19:** Result yielded by running the circuit example for 5,000 shots. Binary numbers were converted to decimals for plotting on the horizontal axis. Vertical axis are the times each of the results were observed.

### 5.3 The sound synthesiser

The system generates sounds using a bespoke additive synthesiser. The additive synthesis technique is informed by the theory of Fast Fourier Transform, or FFT (Muller, 2015). It is based on the notion that sounds can be characterised as a sum of sine waves.

Additive synthesis works by deploying a number of sine wave sound generators (e.g., digital oscillators) to produce partials, which are added up to produce the final result. A sine wave is characterised by a frequency value (i.e., speed of the cycle) and an amplitude value (i.e., strength of the signal).

Our synthesiser comprises eight digital oscillators, each of which requires a frequency value in Herts (Hz) and an amplitude, whose value varies between 0 (silence) and 1 (loudest). The outputs are summed and a Hanning function is applied to the result to give a smooth fade in and fade out bell shape to the sound.

In Figure 20, individual partials are represented on the left-hand side of the figure, where a bar on the 'freq' axis (frequency domain) has a certain magnitude on the 'amp' axis (amplitude domain). The spectrum of the resulting sound is represented on the right-hand side; it contains eight partials. A schematic representation of the resulting sound, depicting its bell-like shape is also shown.

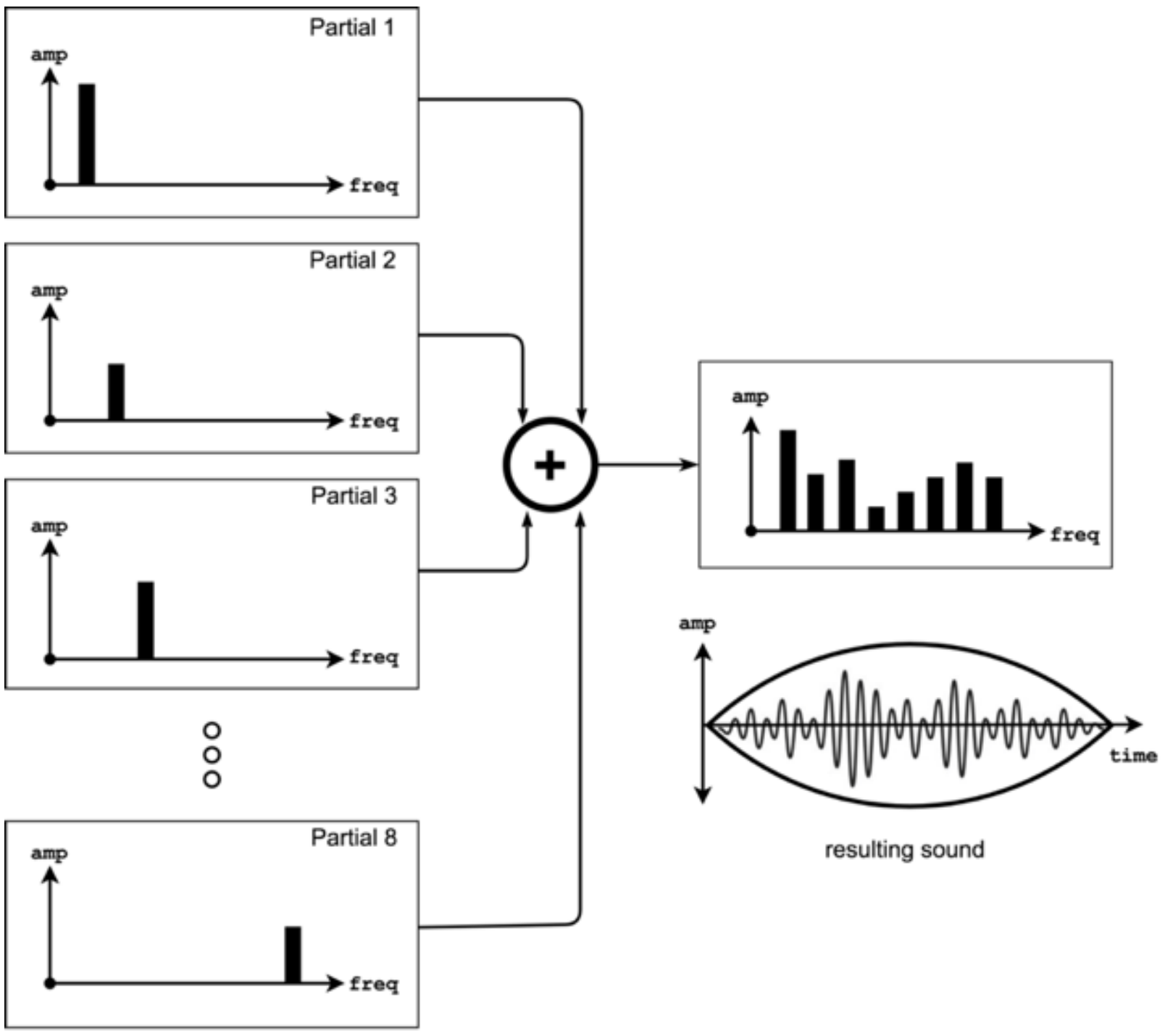

**Figure 20:** Additive sound synthesis works by adding up a number of sine waves.

For every lapse of time the system checks the satisfiability of the respective logic expression, as discussed earlier, and uses the results to activate the oscillators of the synthesiser. Figure 21 delineates how the qubit measurements plotted in Figure 19 are associated with oscillators. Notice that there are as many oscillators as the number of different quantum states that the respective circuit can return: in this case, $2^3 = 8$. That is, each possible output is associated with a different partial of the resulting sound. In additive synthesis, changes to the frequencies and amplitudes of the oscillators modify the timbre of the resulting sound. A schematic representation of this is given in Figure 22.

Each oscillator of the synthesiser is assigned a frequency; for instance: osc 1 = 55.0 Hz, osc 2 = 164.81, osc 3 = 329.63 Hz, and so on. This is fully customisable, and the system can be set to change these frequencies algorithmically. The amplitudes are normalised proportionally to the number of times the respective quantum state was observed after a number of pre-specified shots. Thus, in Figure 21, the first oscillator, whose amplitude is controlled by the number of times the state $|000\rangle$ was observed, will produce a partial that is much quieter than the second oscillator, whose amplitude is given by number of times that state $|001\rangle$ was observed.

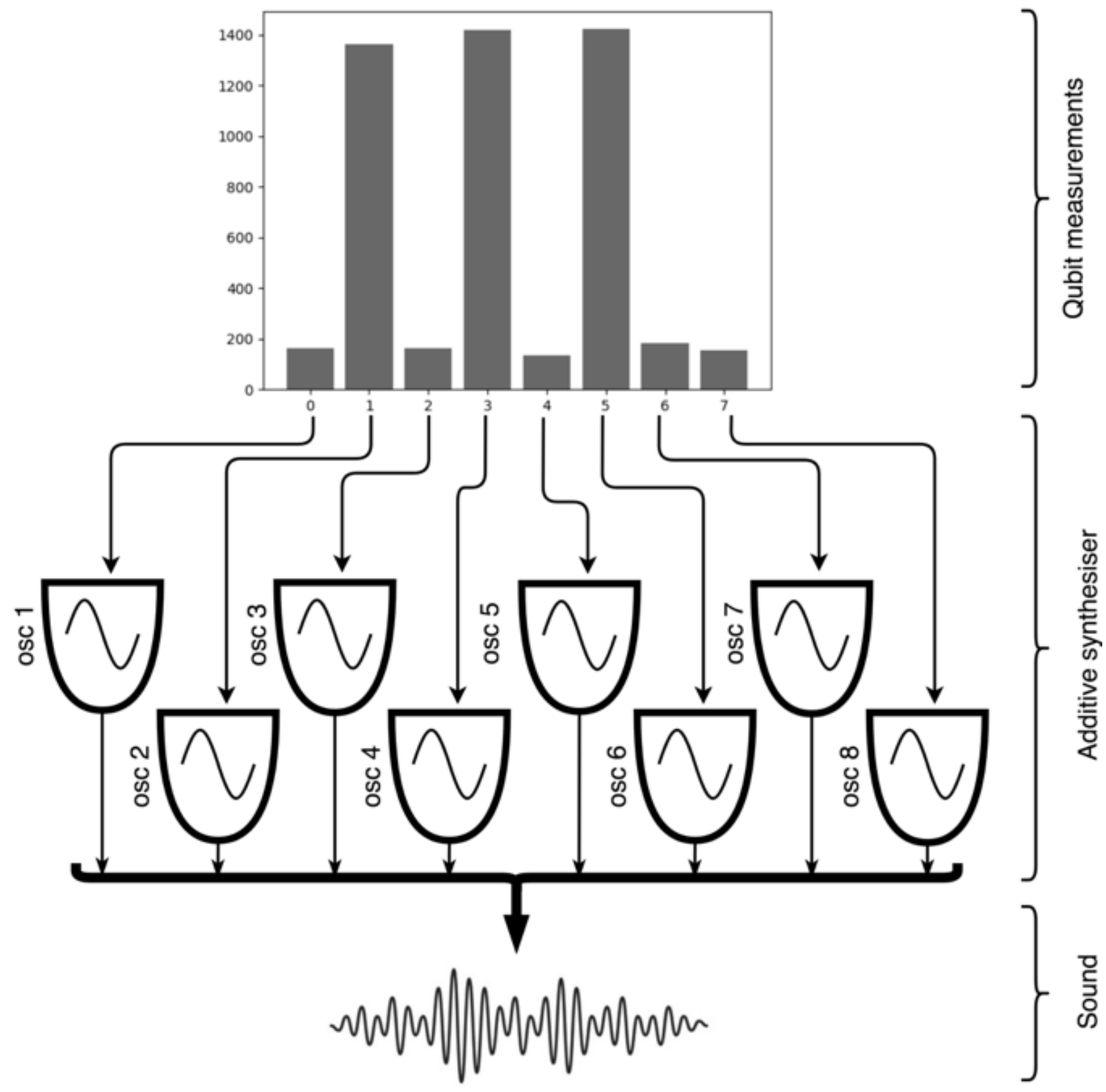

**Figure 21:** From qubit measurements to sound synthesis.

The duration of the time lapses is also customisable. However, it should not be shorter than the window of time that is required to read and analyse EEG data. Our experiments suggest that at least one second is needed to capture sufficient EEG data to associate with a mental state. But in order to perceive differences clearly in the timbre of the sounds, it is suggested to let it play for no less than five seconds. Of course, time can vary and can be defined algorithmically should one wishes to do so.

As a sound is being played, the system processes the EEG for the next time lapse, builds and runs the circuit, and synthesises a new sound immediately after the current one. And the cycle continues for as long as required.

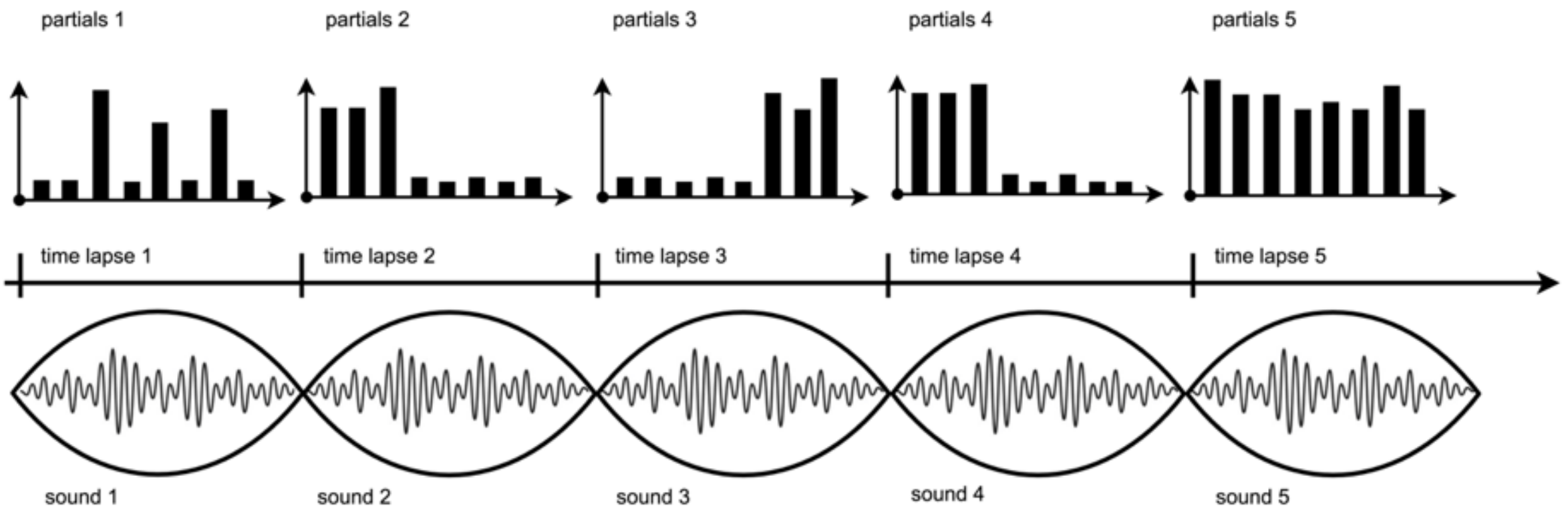

**Figure 22:** Schematic representation of a sequence of five sounds and their respective partials shown above the time lapse line.

## 6 Technical and practical considerations

### 6.1 Reducing outliers

In the example presented in section 5.2, the system spin-marked the quantum states $|001\rangle$, $|011\rangle$ and $|101\rangle$. As shown in Figure 19, it outputted 001, 011 and 101 significantly more times than any of the other possible values. But there are five outliers, whose probability of being observed were nearly 12% of the spin-marked ones. These outliers take place due to the very nature of the amplitude amplification algorithm (Johnston et al., 2019).

The probabilities of the outliers appearing could have been squashed further by repeating the circuit between the initial Hadamard gates and the measurement section a few times. What happens here is that interference can amplify the amplitudes (or 'probabilities') of spin-marked qubits and decrease the amplitudes of all others.

In the case of our example, repeating the circuit three times squashes the outliers to negligible levels, as shown in Figure 23. However, one must exercise caution because the number of repetitions needs to be specified carefully due to the way in which the amplitude amplification technique works. For instance, if the circuit is repeated only twice, then the probabilities of the outliers would increase instead. Moreover, a more important caveat is that such repetitions increases circuit's depth, considerably (i.e., the length of the circuit). On quantum hardware, this escalates significantly the likelihood of errors due to decoherence.

The problem of decoherence poses severe limitations to the quantity of successive gates that can be used in a circuit for a real quantum processor. The higher the number of gates sequenced one after the other, the higher the likelihood of decoherence to occur.

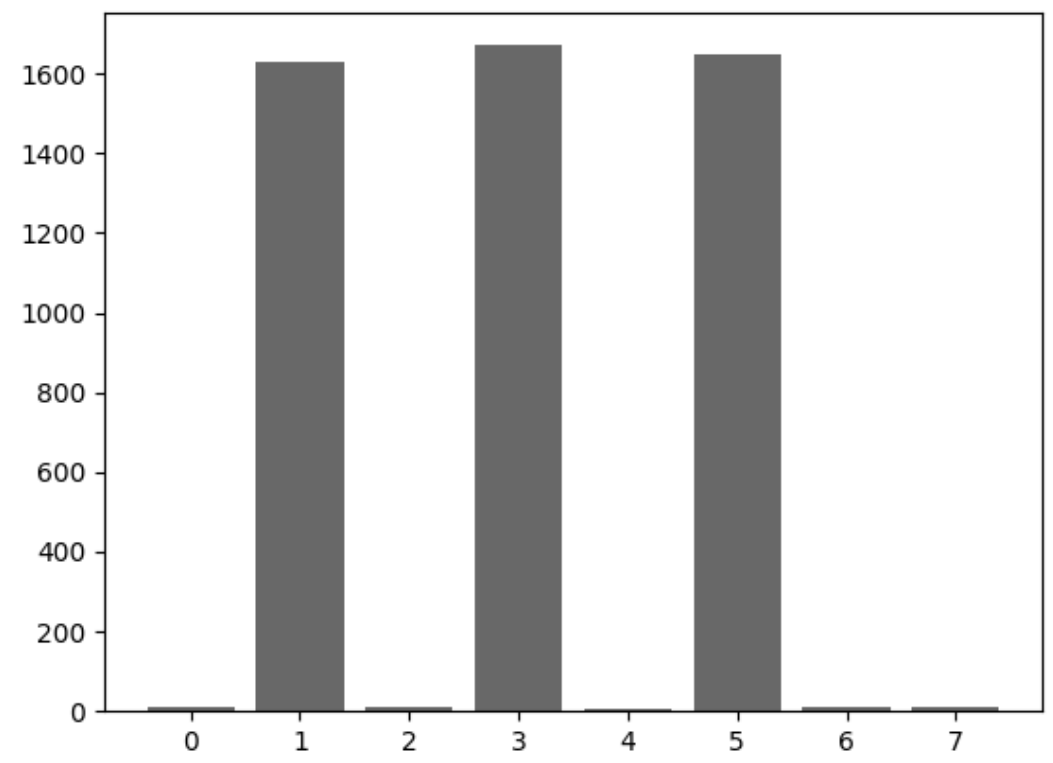

**Figure 23:** Results yielded by running three copies of the circuit in Figure 18 for 5,000 shots on a Rigetti's QVM. Binary numbers were converted to decimals for plotting on the horizontal axis. Vertical axis are the times each of the results were observed.

## 6.2 Running on quantum hardware

At the time of writing, quantum processors struggle to maintain coherence for more than a dozen successive gates involving superposition and entanglement. Currently, coherence is assessed in terms of a few microseconds rather than seconds. However, research is progressing fast to improve this (Cho, 2020). In addition to improving hardware technology, there is much research activity to develop efficient quantum error correction methods.

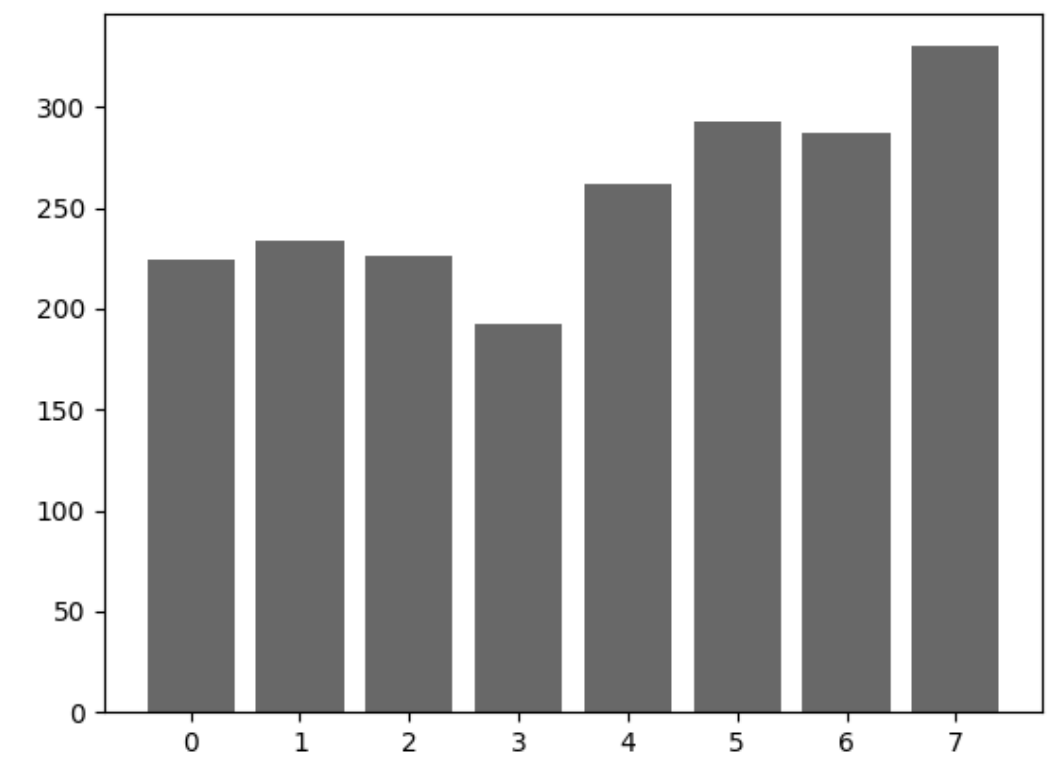

**Figure 24:** Results from running the circuit shown in Figure 18 for 2,048 shots on a Rigetti's Aspen-8 quantum chip. Binary numbers were converted to decimals for plotting on the horizontal axis. Vertical axis are the times each of the results were observed.

In order to illustrate how challenging the problem of decoherence is, let us examine what happened when we ran Code 1 on the state of the art Rigetti's Aspen-8 quantum chip, with no added noise correction algorithm. Aspen-8 affords 3-fold connectivity (i.e., supports 3-qubit gates) and coherence times of approximately 20μs.

Figure 24 plots the outcomes after running the code for 2,048 shots on the quantum chip. The results do not match the ones obtained with the QVM simulator; they are all over the place. This is because the circuit is indeed too deep for this chip.

```
DECLARE ro BIT[3]
H 0
H 1
X 0
X 1
CCNOT 0 1 2
X 0
X 1
X 2
MEASURE 0 ro[0]
MEASURE 1 ro[1]
MEASURE 2 ro[2]
```

**Code 2:** Quil code for the $(\mathrm{A} \vee \mathrm{B})$ portion of the circuit shown in Figure 18.

| $(\mathbf{A} \vee \mathbf{B}) = \|q_2\rangle$ | $\mathbf{B} = \|q_1\rangle$ | $\mathbf{A} = \|q_0\rangle$ | **Outcome** | **Boolean** |
|---|---|---|---|---|
| 0 | 0 | 0 | 000 = 0 | True |
| 0 | 0 | 1 | 001 =1 | False |
| 0 | 1 | 0 | 010 = 2 | False |
| 0 | 1 | 1 | 011 = 3 | False |
| 1 | 0 | 0 | 100 = 4 | False |
| 1 | 0 | 1 | 101 = 5 | True |
| 1 | 1 | 0 | 110 = 6 | True |
| 1 | 1 | 1 | 111 = 7 | True |

**Table 4**: The outcomes that satisfy the logical clause $(\mathrm{A} \vee \mathrm{B})$**.**

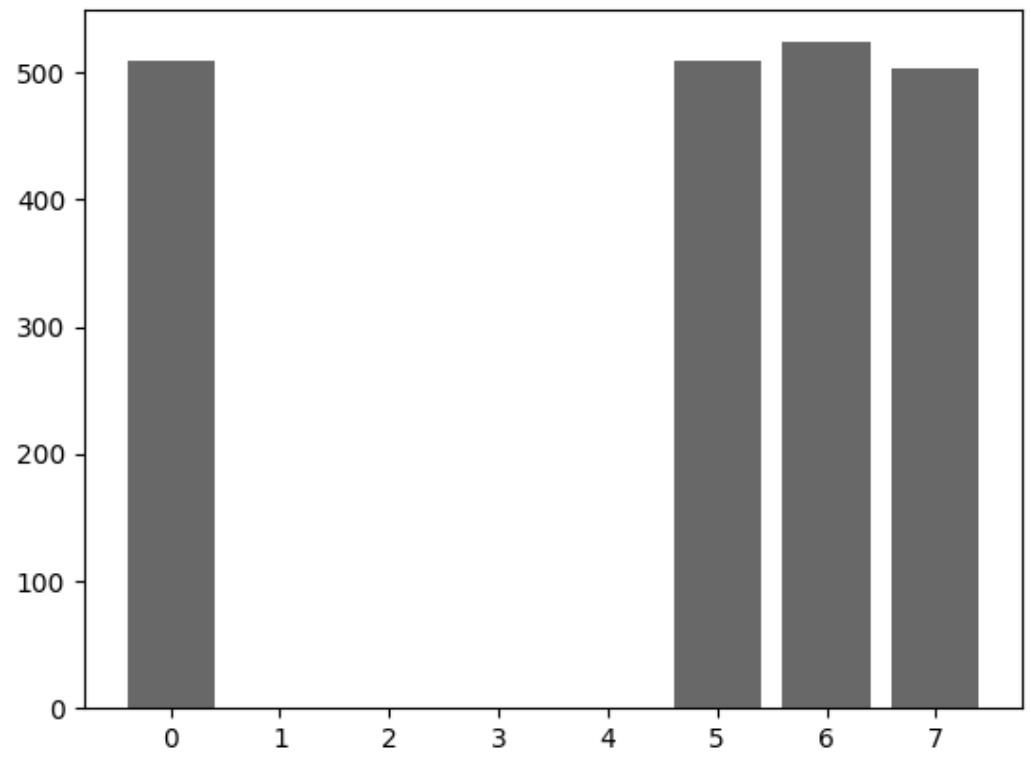

**Figure 25:** Results from running the circuit shown in Code 2 for 2,048 shots on Rigetti's Quantum Virtual Machine (QVM). Binary numbers were converted to decimals for plotting on the horizontal axis. Vertical axis are the times each of the results were observed.

In truth, even the circuit for the first logic clause $(\mathrm{A} \vee \mathrm{B})$ is problematic. The Quil code to examine the satisfiability of this simple clause is shown in Code 2. Table 4 shows the outcomes that satisfy this clause. Thus, one can predict that each time we run Code 2, it would output either 000, 101, 110, 111. Indeed, Figure 25 shows the times that each of the possible outcomes were observed after running the circuit for 2,048 shots on Rigetti's QVM. They match the prediction perfectly. Yet, the outcomes from running exactly the same code on Aspen-8 are not accurate (Figure 26). Why is this so?

Fundamentally, quantum computing programming languages are built upon a handful of universal gates that physically act on quantum chips. Rigetti's Aspern-8 is a superconducting quantum chip, which can enact the following basic gates: Rx($\vartheta$), Rz($\vartheta$), CZ and XY($\vartheta$)[8]. These are referred to as Quil's native gates. The measurements are done natively in the computational basis (z-axis). All other standard Quil gates (H, X, Z, Y, CNOT, and so on) are built using native gates. Thus, before compilation, a standard Quil code needs to be transpiled[9] to native Quil. Code 3 shows the transpilation from Code 2.

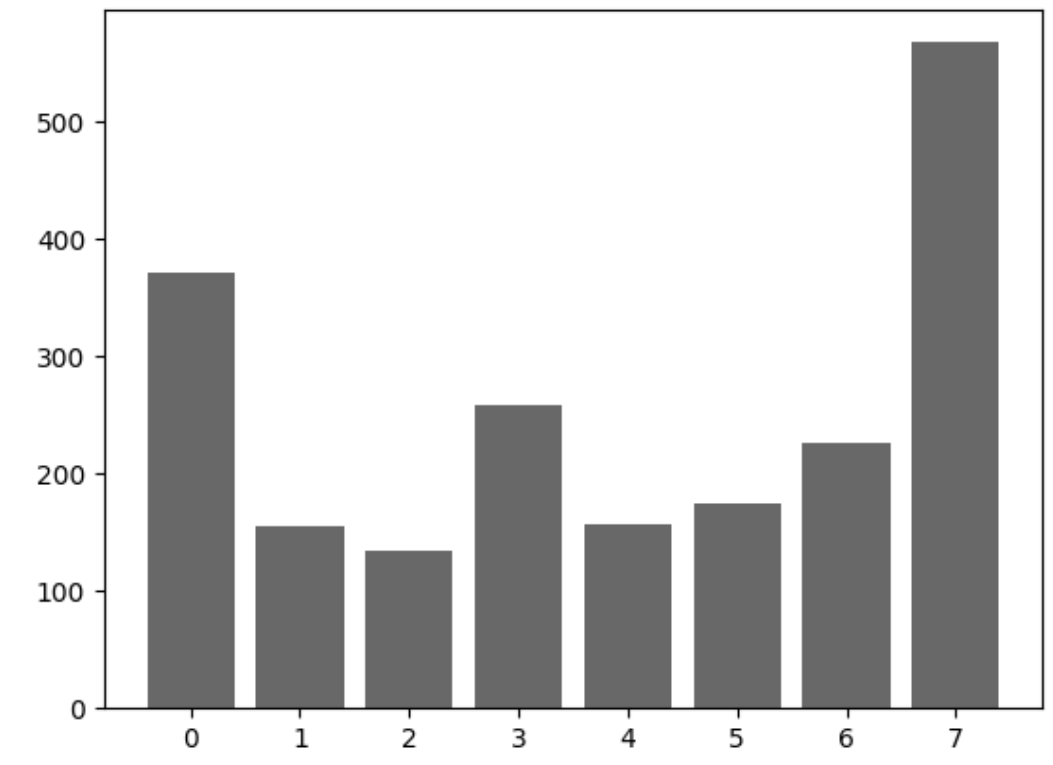

**Figure 26:** Results from running the circuit shown in Code 2 for 2048 shots for 2048 trials on Rigetti's Aspen-8 quantum chip. Binary numbers were converted to decimals for plotting on the horizontal axis. Vertical axis are the times each of the results were observed.

Even though the circuit shown in Code 2 is 12 lines long and uses only eight standard gates, its transpilation results in 39 lines of code with as many as 35 native Quil gates. Unfortunately, this is too deep for Aspen-8. The real culprit here is the CCNOT, Toffoli gate. This gate is notoriously expensive in native currency.

Similar tests on an IBM's quantum computing resources can be found in Appendix II.

```
DECLARE ro BIT[3]
RX(pi) 5
RZ(pi/2) 6
RX(-pi/2) 6
CZ 5 6
RZ(-pi/2) 5
RX(pi/2) 5
RZ(pi/4) 5
RX(-pi/2) 5
RZ(pi/2) 7
RX(-pi/2) 7
CZ 5 7
RX(pi/2) 5
RZ(-pi/4) 5
RX(-pi/2) 5
CZ 5 6
RX(pi/2) 5
RZ(pi/4) 5
RX(-pi/2) 5
CZ 7 5
RX(pi/2) 5
RZ(3*pi/4) 5
RX(pi/2) 5
RZ(-pi/2) 5
MEASURE 5 ro[2]
RZ(-3*pi/4) 6
RX(pi/2) 6
CZ 6 7
RX(pi/2) 6
RZ(pi/4) 6
RX(-pi/2) 6
CZ 6 7
RX(pi/2) 6
RZ(-pi/2) 6
MEASURE 6 ro[1]
RZ(-pi/4) 7
RX(pi) 7
MEASURE 7 ro[0]
HALT
```

**Code 3:** The transpiled native version of the code shown in Code 2.

[8] XY($\vartheta$) produces a parameterized iSWAP gate, which is not priority for discussion on this chapter. For more details see (Abrams et al., 2019).

[9] A transpiler is a program that translates a piece of code into another at the same level of abstraction. It is different from a compiler whose output is in a lower level of abstraction than the input.

## 6.3 Quantum advantage

The expressions in section 5.1 are limited to three logic variables. One may not even need a computer to check their satisfiability, even less so a quantum computer. Should the clauses have entailed a unique variable for each electrode, then the task of checking the expressions' satisfiability would have been harder. But still, not hard enough to justify the need for a quantum computer.

However, if the expressions encompassed the whole set of 20 electrodes shown in Figure 3, each of which corresponding to a unique logic variable, then the satisfiability problem would become considerably harder. In this case a quantum computer could well be advantageous.

The amplitude amplification technique is a core component of the so-called Grover's algorithm. Introduced by Grover (1996), this algorithm uses amplitude amplification to search for an element in an unstructured set of $N$ elements. A brute-force classic algorithm would scan all elements in the set until it finds the one that is sought after. In the worst-case scenario, the element in question could have been the last one to be checked, which means that the algorithm would have made $N$ queries to find it. Provided that a given problem can be encoded efficiently in terms of qubits, Grover's algorithm would be able find a solution with $\sqrt{N}$ queries. Thus, Grover's algorithm provides a quadratic speedup. This benchmarking also applies for logic satisfiability problems.

Ignoring for the moment the format of the logical expressions discussed in section 5.1, let us suppose that a system would need to verify if the values A = True, B = True and C = False satisfy a given logical expression. In this case, the three Boolean variables amount to eight possible combinations for A, B, and C; that is, $2^3 = 8$. Therefore, a brute-force algorithm would need to make up to eight checks to get the answer. That is, the system would have to run the algorithm up to eight times. Conversely, Grover's algorithm could solve this with $\sqrt{8} = 2.8$ runs. Indeed, recall that the convincing results shown in Figure 23 were obtained by running three copies of the circuit.

The difference in performance escalates significantly as the number of logic variables increases. For instance, 20 variables would require up to $2^{20} = 1{,}048{,}576$ checks classically. Whereas Grover's algorithms would require $\sqrt{2^{20}}$ = 1,024 runs.

Brushing aside any thorough comparison between the processing clocks of classical and quantum hardware, a quantum computer running Grover's algorithms would certainly outperform a classical computer running brute-force search. However, the slightly disappointing news is that the number of qubits needed to implement large Grover's-like circuits is prohibitive the present times.

Nevertheless, despite the limited capacity of current quantum hardware technology, the quantum computing research community remains optimistic. The industry is aiming at producing quantum chips with over 1,000 qubits by 2023 (Cho, 2020).

## 7 Concluding remarks

This chapter presented an approach to interfacing the brain and a quantum computer. Central to this approach is a method to encode brain activity as logic expressions representing states of mind. These expressions are associated with commands to control systems and/or machines with brain signals.

In addition to building brain-computer interfaces, the chapter put forward the notion of a logic of the mind as a means to study neural correlates of the mind and brain functioning, with potential benefits to medicine; e.g., for diagnosing brain disorder. The EEG is an important physiological signal for diagnostics (Tatum IV, 2008). It is used to detect traces of epilepsy, dementia, cancer, inflammation, sleep disorders, and more. For instance, Clarke et al. (2013) reported that persistent excess of beta rhythms at the frontal regions of the brain can indicate hyperactivity disorder. It would be straightforward to encode this as a logic of the mind expression.

The EEG is currently the best signal for BCI systems. However, it is not necessarily the best for research into understanding brain functioning and mental correlates. For instance, functional magnetic resonance imaging (fMRI) measures the minuscule variations in blood flow that occurs with brain activity. It is used to determine which parts of the brain are more active than others when handling particular tasks. It has higher spatial resolution than EEG has, and it can detect information deep inside the brain.

The logic of the mind requires the processing of large Boolean expressions, which are computationally very demanding. Even more so if we were to use fMRI scanning. Today's average classical computer would take hours, if not days, to evaluate a handful of expressions. Hence the rationale for using quantum computers.

At the time of writing, quantum computing hardware is of limited capacity for realistic applications. But once fully developed, quantum computers promise to be considerably faster than their classical counterparts to run certain types of algorithms, such as the one introduced in this chapter. In the meantime, simulations on virtual quantum machines enable research and development of quantum algorithms, which would eventually run optimally when robust quantum hardware is available.

By way of demonstration, an application was presented, which renders the logic of the mind into sounds. We introduced a method to sonify quantum measurements obtained from evaluating logic expressions. The demonstration illustrates a slightly different approach to BCI, where states of mind are mapped directly onto sound, instead of control commands. Effectively, the system can be thought of as a novel musical instrument.

Incidentally, the sound synthesis technique introduced above is a contribution to the field quantum computing on its own right, in the sense that it provides a method to represent the wavefunction of a quantum system auditorily.

## Acknowledgements

The author would like to thank Mathew Wilson and James Hefford of the Department of Computer Science at University of Oxford for reviewing the quantum computing content of this chapter. Also, many thanks to Palaniappan Ramaswamy of the School of Computing at University of Kent for scrutinising the BCI content. Their meticulous comments and suggestions contributed significantly to the clarity and rigour of this work.

## Appendix 1: Examples from the BCI system

Below are four examples of results taken from a run of the BCI system presented in this chapter, using Rigetti's QVM. Different results produce variations in the timbre of the sounds.

Recordings of the respective sounds, and other examples, are available online at SoundClick: https://soundclick.com/LogicoftheMind

### A1 Example 1

This is an example where the logic expression is unsatisfiable.

| **Expression:** $(\neg C \vee B) \wedge (C \vee A) \wedge (\neg C \vee B)$ | | |
|---|---|---|
| **Clause 1:** $(\neg C \ \vee B)$ | **Clause 2:** $(C \ \vee A)$ | **Clause 3:** $(\neg C \ \vee B)$ |
| **Term 1:**<br>Selected electrode = O1<br>Logic variable = C<br>Frequency = 13.1649 Hz | **Term 1:**<br>Selected electrode = Oz<br>Logic variable = C<br>Frequency = 31.1541 Hz | **Term 1:**<br>Selected electrode = O2<br>Logic variable = C<br>Frequency = 8.22338 Hz |
| **Term 2:**<br>Selected electrode = T3<br>Logic variable = B<br>Frequency = 24.7721 Hz | **Term 2:**<br>Selected electrode = Fz<br>Logic variable = A<br>Frequency = 31.5992 Hz | **Term 2:**<br>Selected electrode = T4<br>Logic variable = B<br>Frequency = 27.0611 Hz |

**Table A1.1:** The EEG analysis and generated expression for Example 1.

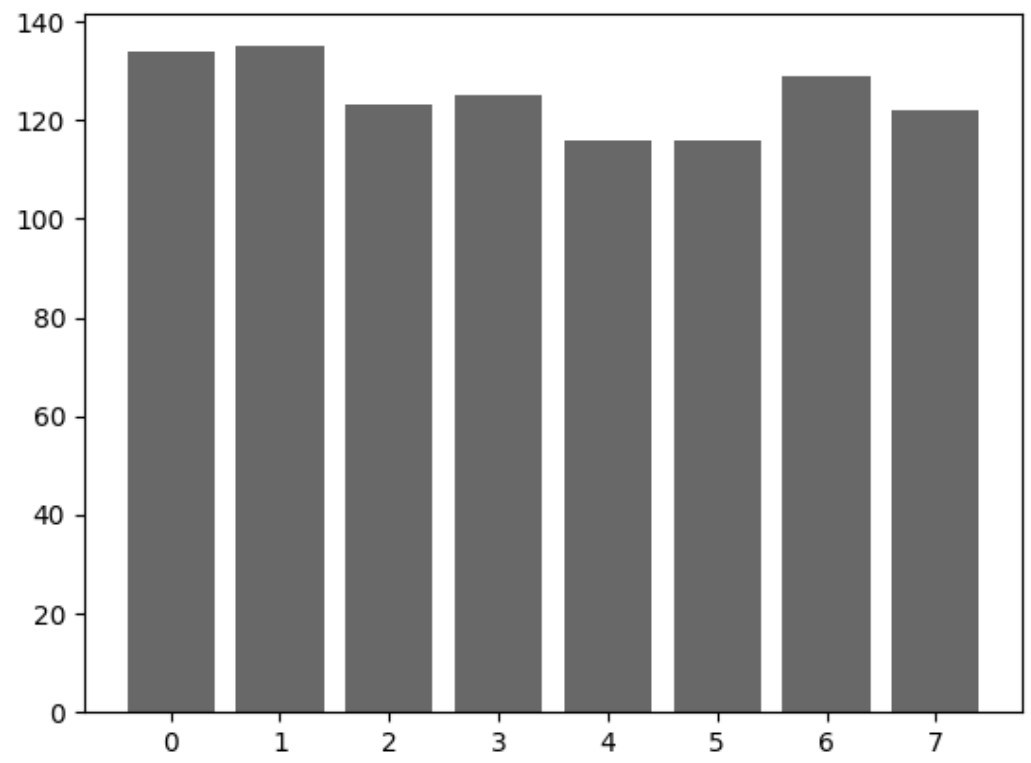

**Figure A1.1:** Results from running the circuit for Example 1 for 1,000 times. Binary numbers were converted to decimals for plotting on the horizontal axis. Vertical axis are the times each of the results were observed.

```
H 0
H 1
H 2
X 2
X 2
X 1
CCNOT 2 1 3
X 2
X 1
X 3
X 2
X 2
X 0
CCNOT 2 0 4
X 2
X 0
X 4
X 2
X 2
X 1
CCNOT 2 1 5
X 2
X 1
X 5
X 2
CONTROLLED CONTROLLED Z 5 4 3
X 2
X 1
X 5
X 2
CCNOT 2 1 5
X 2
X 2
X 1
X 2
X 0
X 4
CCNOT 2 0 4
X 2
X 0
X 2
X 1
X 3
X 2
CCNOT 2 1 3
X 2
X 2
X 1
H 0
H 1
H 2
X 0
X 1
X 2
CONTROLLED CONTROLLED Z 0 1 2
X 0
X 1
X 2
H 0
H 1
H 2
DECLARE ro BIT[3]
MEASURE 0 ro[0]
MEASURE 1 ro[1]
MEASURE 2 ro[2]
```

**Code A1.1:** Generated Quil code for Example 1.

## A2 Example 2

| **Expression:** $(B \vee A) \wedge (C \vee A) \wedge ((\neg A \ \vee \neg C)$ | | |
|---|---|---|
| **Clause 1:** $(B \vee A)$ | **Clause 2:** $(C \vee A)$ | **Clause 3:** $(\neg A \ \vee \neg C)$ |
| **Term 1:**<br>Selected electrode = T3<br>Logic variable = B<br>Frequency = 20.6042 Hz | **Term 1:**<br>Selected electrode = Oz<br>Logic variable = C<br>Frequency = 18.7471 Hz | **Term 1:**<br>Selected electrode = Fp2<br>Logic variable = A<br>Frequency = 8.0119 Hz |
| **Term 2:**<br>Selected electrode = Fp1<br>Logic variable = A<br>Frequency = 21.2267 Hz | **Term 2:**<br>Selected electrode = Fz<br>Logic variable = A<br>Frequency = 32.5744 Hz | **Term 2:**<br>Selected electrode = O2<br>Logic variable = C<br>Frequency = 10.3202 Hz |

**Table A1.2:** The EEG analysis and generated expression for Example 2.

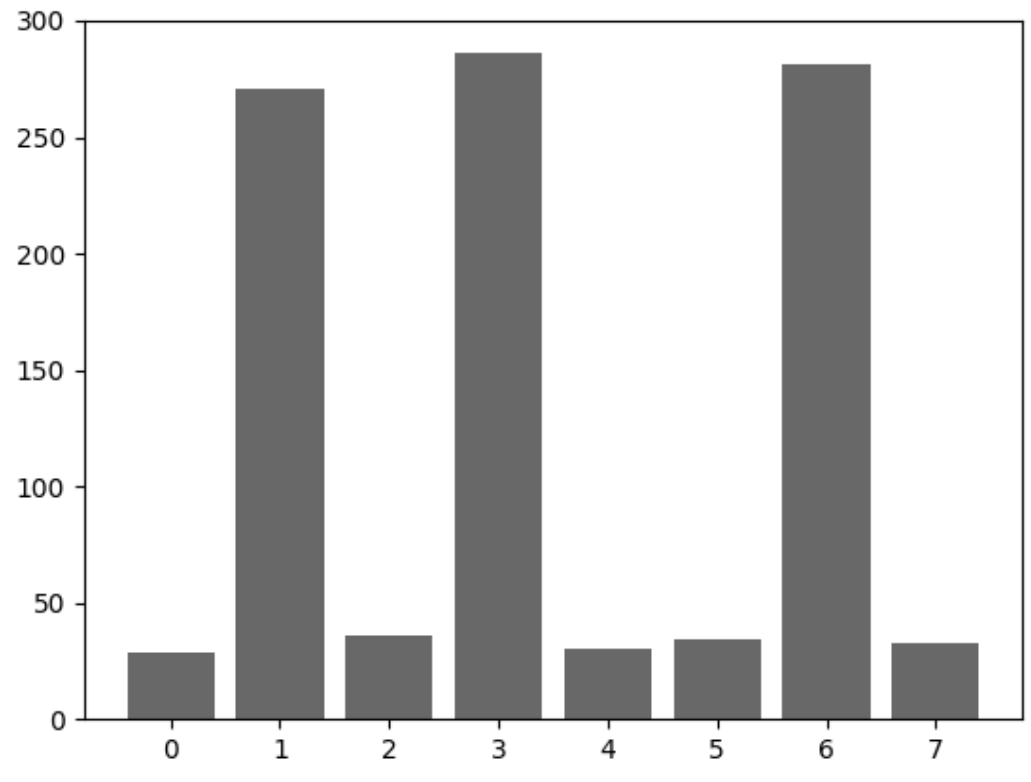

**Figure A1.2:** Results from running the circuit for Example 2 for 1,000 times. Binary numbers were converted to decimals for plotting on the horizontal axis. Vertical axis are the times each of the results were observed.

```
H 0
H 1
H 2
X 1
X 0
CCNOT 1 0 3
X 1
X 0
X 3
X 2
X 0
CCNOT 2 0 4
X 2
X 0
X 4
X 0
X 2
X 0
X 2
CCNOT 0 2 5
X 0
X 2
X 5
X 0
X 2

CONTROLLED CONTROLLED Z 5 4 3
X 0
X 2
X 5
X 0
X 2
CCNOT 0 2 5
X 0
X 2
X 0
X 2
X 2
X 0
X 4
CCNOT 2 0 4
X 2
X 0
X 1
X 0
X 3
CCNOT 1 0 3
X 1
X 0
H 0

H 1
H 2
X 0
X 1
X 2
CONTROLLED CONTROLLED Z 0 1 2
X 0
X 1
X 2
H 0
H 1
H 2
DECLARE ro BIT[3]
MEASURE 0 ro[0]
MEASURE 1 ro[1]
MEASURE 2 ro[2]
```

**Code A1.2:** Generated Quil code for Example 2.

## A3 Example 3

| **Expression:** $(\neg C \vee A) \wedge (C \vee A) \wedge (\neg B \vee \neg A)$ | | |
|---|---|---|
| **Clause 1:** $(\neg C \vee A)$ | **Clause 2:** $(C \vee A)$ | **Clause 3:** $(\neg B \vee \neg A)$ |
| **Term 1:**<br>Selected electrode = O1<br>Logic variable = C<br>Frequency = 13.7849 Hz | **Term 1:**<br>Selected electrode = Oz<br>Logic variable = C<br>Frequency = 17.5519 Hz | **Term 1:**<br>Selected electrode = T4<br>Logic variable = B<br>Frequency = 12.6194 Hz |
| **Term 2:**<br>Selected electrode = Fp1<br>Logic variable = A<br>Frequency = 23.9491 Hz | **Term 2:**<br>Selected electrode = Fz<br>Logic variable = A<br>Frequency = 18.6791 Hz | **Term 2:**<br>Selected electrode = Fp2<br>Logic variable = A<br>Frequency = 13.1322 Hz |

**Table A1.3:** The EEG analysis and generated expression for Example 3.

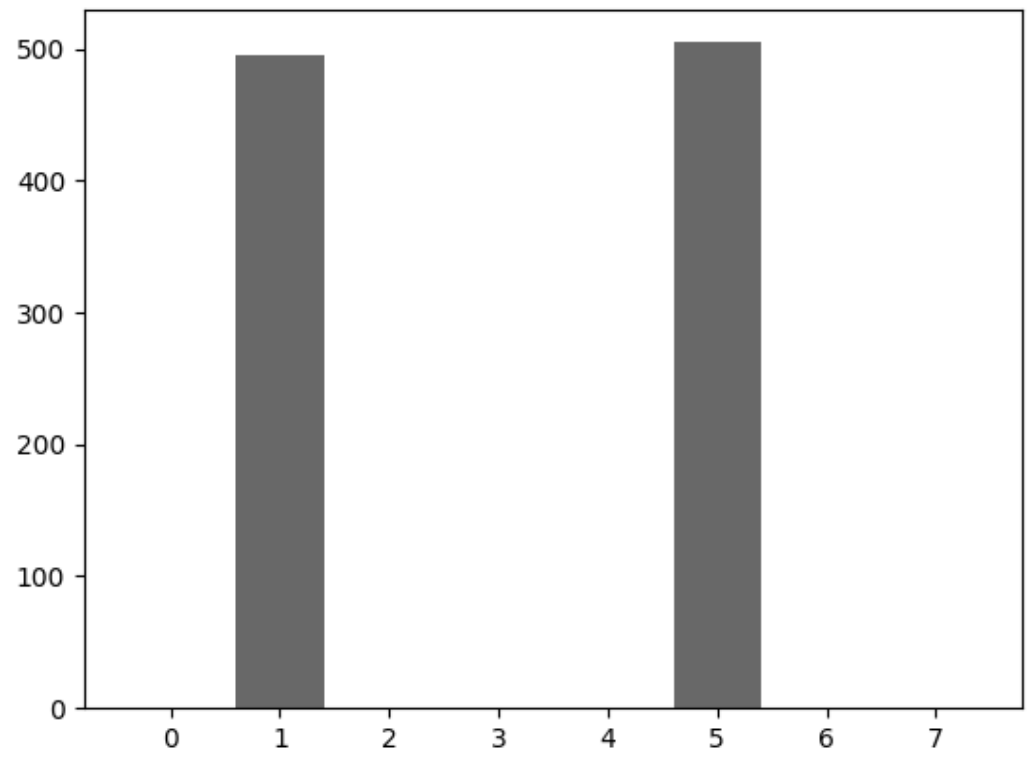

**Figure A1.3:** Results from running the circuit for Example 3 for 1,000 times. Binary numbers were converted to decimals for plotting on the horizontal axis. Vertical axis are the times each of the results were observed.

```
H 0
H 1
H 2
X 2
X 2
X 0
CCNOT 2 0 3
X 2
X 0
X 3
X 2
X 2
X 0
CCNOT 2 0 4
X 2
X 0
X 4
X 1
X 0
X 1
X 0
CCNOT 1 0 5
X 1
X 0
X 5
X 1
X 0
CONTROLLED CONTROLLED Z 5 4 3
X 1
X 0
X 5
X 1
X 0
CCNOT 1 0 5
X 1
X 0
X 1
X 0
X 2
X 0
X 4
CCNOT 2 0 4
X 2
X 0
X 2
X 0
X 3
X 2
CCNOT 2 0 3
X 2
X 2
X 0
H 0
H 1
H 2
X 0
X 1
X 2
CONTROLLED CONTROLLED Z 0 1 2
X 0
X 1
X 2
H 0
H 1
H 2
DECLARE ro BIT[3]
MEASURE 0 ro[0]
MEASURE 1 ro[1]
MEASURE 2 ro[2]
```

**Code A1.3:** Generated Quil code for Example 3.

## A4 Example 4

| **Expression:** $(A \vee B) \wedge (B \vee A) \wedge (B \vee C)$ | | |
|---|---|---|
| **Clause 1:** $(A \vee B)$ | **Clause 2:** $(B \vee A)$ | **Clause 3:** $(B \vee C)$ |
| **Term 1:**<br>Selected electrode = Fp1<br>Logic variable = A<br>Frequency = 15.0409 Hz | **Term 1:**<br>Selected electrode = Cz<br>Logic variable = B<br>Frequency = 30.1681 Hz | **Term 1:**<br>Selected electrode = T4<br>Logic variable = B<br>Frequency = 18.4086 Hz |
| **Term 2:**<br>Selected electrode = T3<br>Logic variable = B<br>Frequency = 18.6357 Hz | **Term 2:**<br>Selected electrode = Fz<br>Logic variable = A<br>Frequency = 32.1824 Hz | **Term 2:**<br>Selected electrode = O2<br>Logic variable = C<br>Frequency = 19.1024 Hz |

**Table A1.4:** The EEG analysis and generated expression for Example 4.

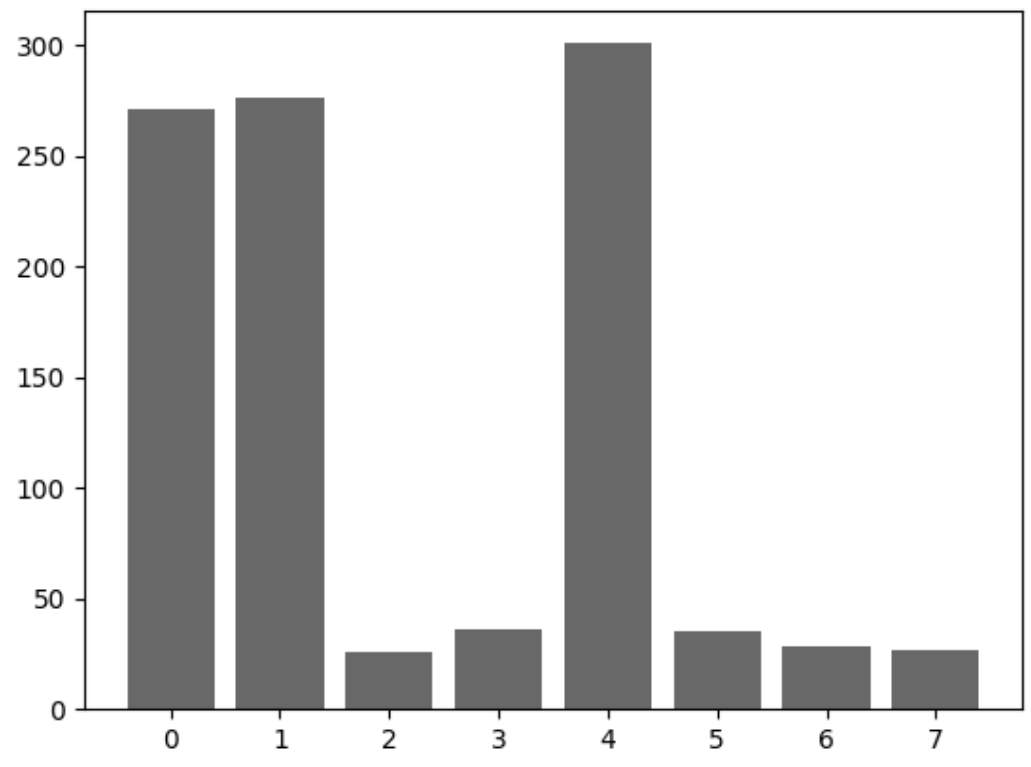

**Figure A1.4:** Results from running the circuit for Example 4 for 1,000 times. Binary numbers were converted to decimals for plotting on the horizontal axis. Vertical axis are the times each of the results were observed.

```
H 0
H 1
H 2
X 0
X 1
CCNOT 0 1 3
X 0
X 1
X 3
X 1
X 0
CCNOT 1 0 4
X 1
X 0
X 4
X 1
X 2
CCNOT 1 2 5
X 1
X 2
X 5
CONTROLLED CONTROLLED Z 5 4 3
X 1
X 2
X 5
CCNOT 1 2 5
X 1
X 2
X 1
X 0
X 4
CCNOT 1 0 4
X 1
X 0
X 0
X 1
X 3
X 1
X 2
X 5
CCNOT 2 0 3
X 2
X 2
X 0
H 0
H 1
H 2
X 0
X 1
X 2
CONTROLLED CONTROLLED Z 0 1 2
X 0
X 1
X 2
H 0
H 1
H 2
DECLARE ro BIT[3]
MEASURE 0 ro[0]
MEASURE 1 ro[1]
MEASURE 2 ro[2]
```

**Code A1.4:** Generated Quil code for Example 4.

## Appendix 2: Study using IBM Quantum Computing Resources

Figure A2.1 shows the results from running the excerpt corresponding to the logic clause $(A \lor B)$ on an IBM quantum computer simulator, for 2,048 shots. The Quil code shown in Code 2, in section 6.2, was translated into the IBM's OpenQASM language, as shown in Code A2.1.

```
qreg q[3];
creg c[3];
h q[0];
h q[1];
x q[0];
x q[1];
ccx q[0],q[1],q[2];
x q[0];
x q[1];
x q[2];
measure q[0] -> c[0];
measure q[1] -> c[1];
measure q[2] -> c[2];
```

**Code A2.1:** OpenQASM version of the Quil code shown in Code 2.

Not surprisingly, the results shown in Figure A2.1 are identical to those obtained with Rigetti's QMV, as shown in Figure 25.

Furthermore, the outcomes from running the exact OpenQASM code on IBM's Santiago processor (*ibmq_santiago* v1.1.1) is shown Figure A2.2. Although there were errors, in overall the results are comparable to those produced by the simulator. IBM's Santiago processor proved to be more resilient than Rigetti's Aspen-8 in this case.

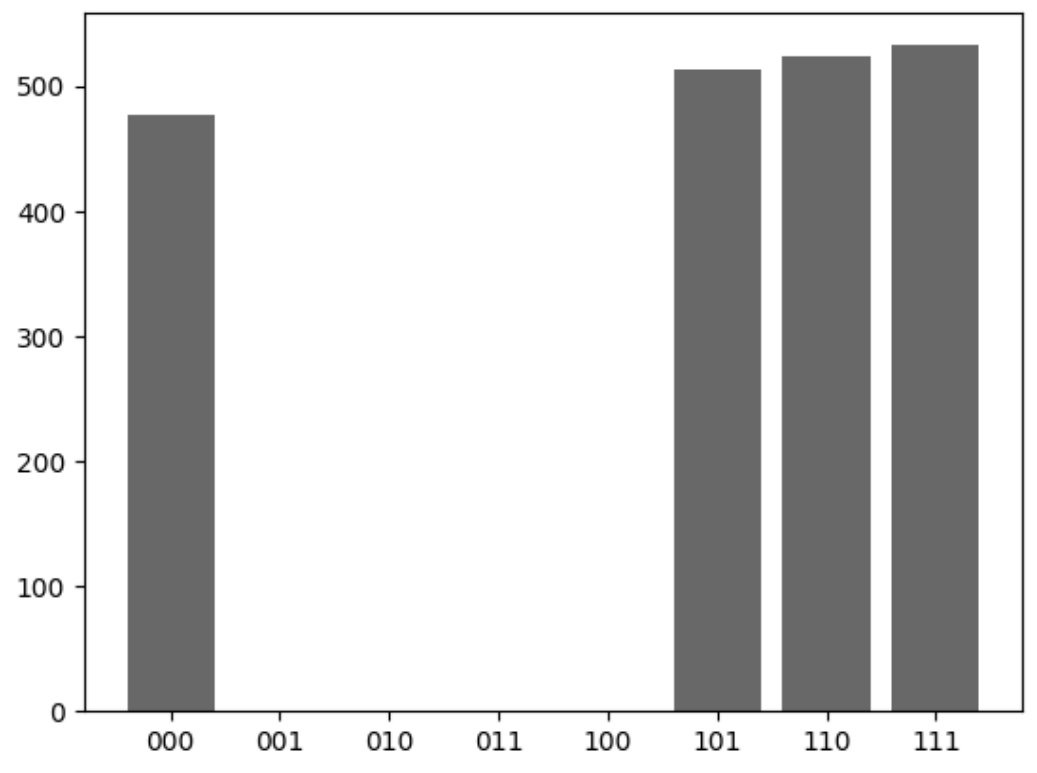

**Figure A2.1:** Results from running the circuit for the logic clause $(A \vee B)$ for 2,048 trials on IBM's simulator.

However, it should be noted the programming for both machines were kept simple, with no error correction and no a priori hardware calibration. Aspen-8 might have performed just as well with appropriate calibration and noise mitigation procedures. A detailed technical discussion on the idiosyncrasies of these machines, calibration, error mitigation strategies, and so on, is beyond the scope of this chapter.

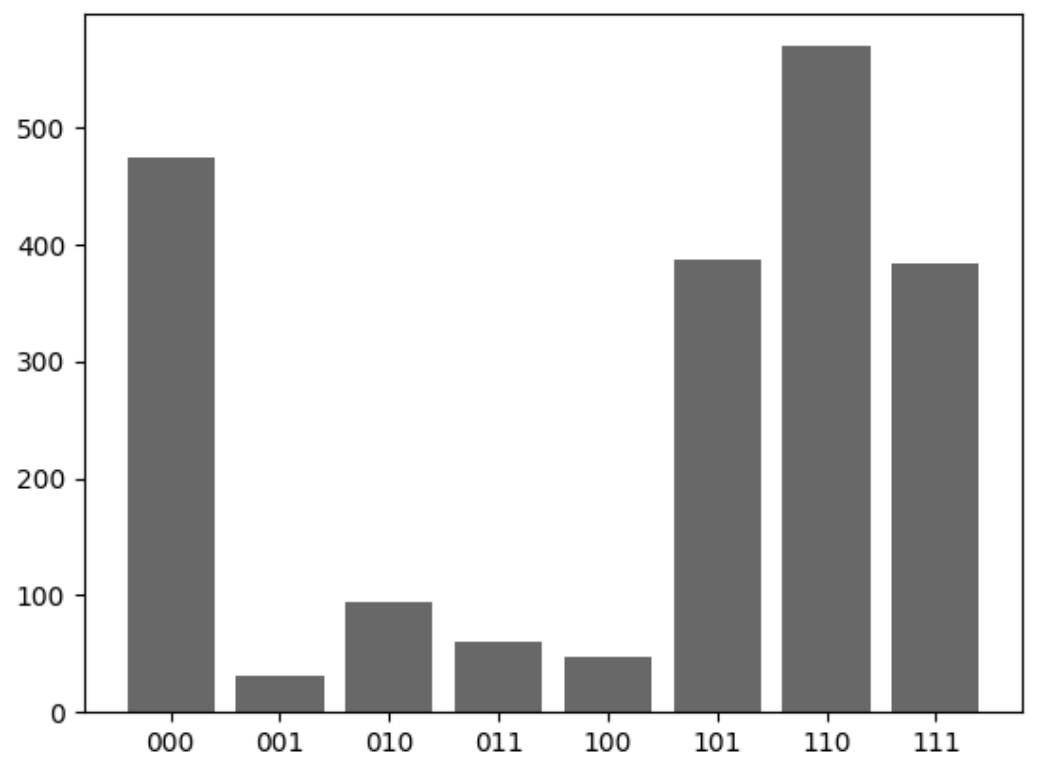

**Figure A2.2:** Results from running the circuit for the logic clause $(A \vee B)$ for 2,048 trials on IBM's Santiago processor.

In order to assess how well IBM's hardware would fare with a deeper circuit, we added a third logic variable and a phase-logic AND operation to implement the expression $((A \vee B) \wedge C)$, as shown in Figure A2.3.

Considering that qubit strings are enumerated from the right end of the string to the left (i.e., $|CBA\rangle$), the expression $((A \vee B) \wedge C$ is satisfied when:

Case 1, $|101\rangle$: A = True, B = False and C = True
Case 2, $|110\rangle$: A = False, B = True and C =True
Case 3, $|111\rangle$: A = True, B = True and C = True

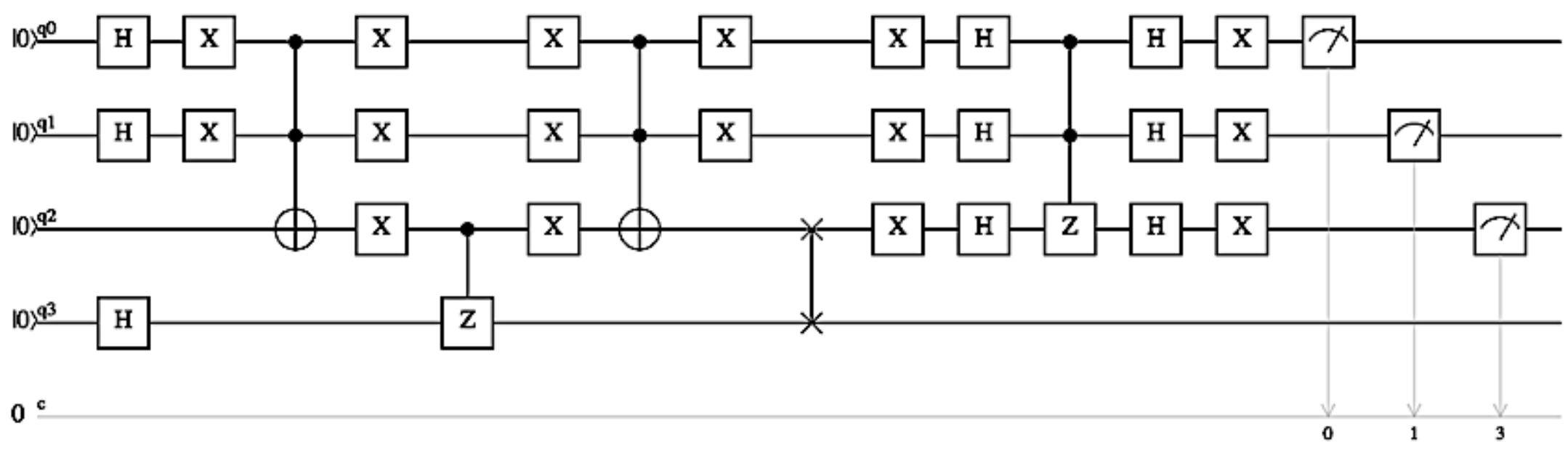

**Figure A2.3:** Quantum circuit for the expression $((A \vee B) \wedge C)$.

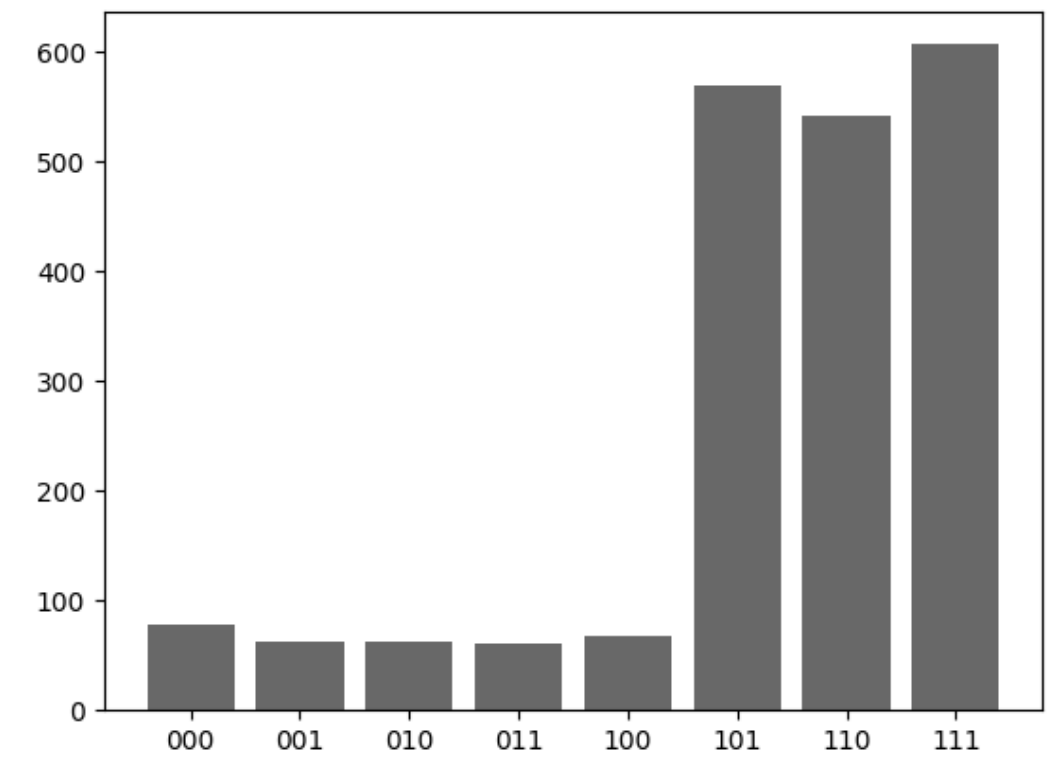

**Figure A2.4:** Results from running the circuit in Figure A.3 for 2,048 trials on IBM's simulator.

Figure A2.4 shows the outcomes from running the circuit on the IBM quantum computer simulator, for 2,048 shots. Not surprisingly, the measurements are fairly accurate. Yet, the outcomes from running exactly the same code on IBM's Santiago processor (*ibmq_santiago* v1.1.1) are not accurate (Figure A2.5). The transpiled code comprised 75 gates; the circuit is too deep for this processor.

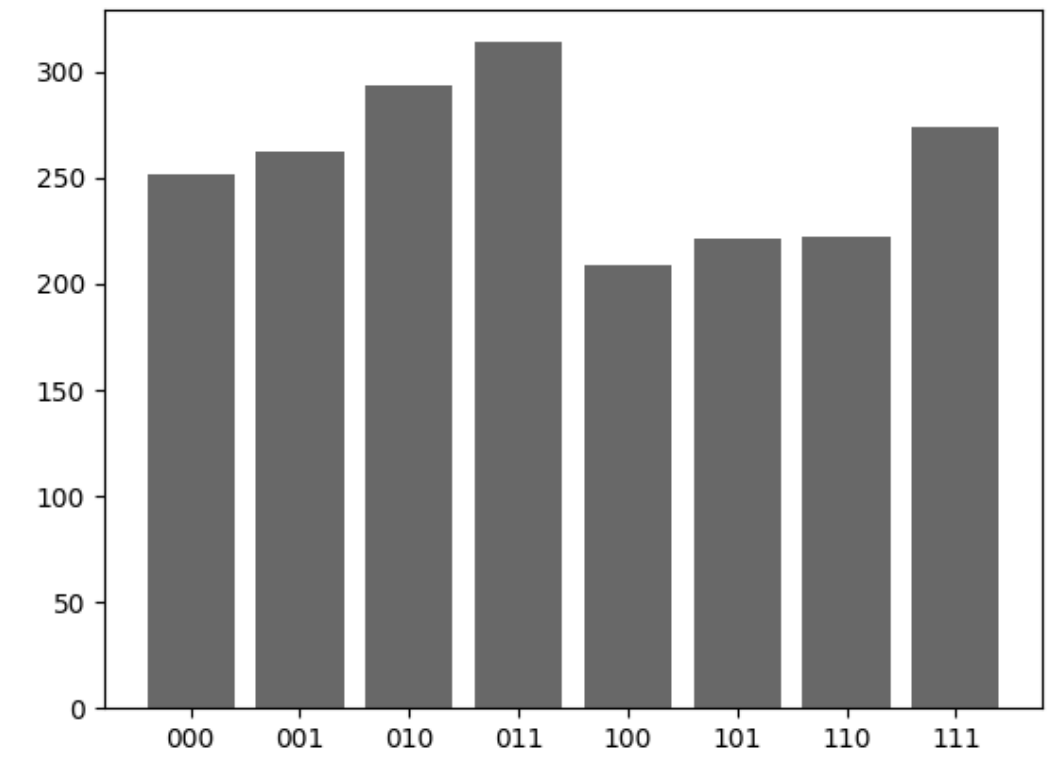

**Figure A2.5:** Results from running the circuit in Figure A.3 for 2048 trials on IBM's Santiago processor.